\documentclass[12pt]{iopart}

\usepackage{graphicx}
\usepackage{dcolumn}
\usepackage{bm}
\usepackage[english]{babel}
\usepackage[latin1]{inputenc} 
\usepackage{graphicx}
\usepackage{epstopdf}
\usepackage{amsfonts}
\usepackage{color}
\usepackage{epsfig}
\usepackage{mathrsfs}
\usepackage{iopams}  
\usepackage{subfigure}

\usepackage{framed}
\usepackage{lipsum}
\usepackage[svgnames]{xcolor}
\definecolor{shadecolor}{named}{LightGrey}

\newcommand{\lto}[1]{\longrightarrow#1}

\renewcommand{\(}{\left(}
\renewcommand{\)}{\right)}
\renewcommand{\[}{\left[}
\renewcommand{\]}{\right]}

\newcommand{\text}[1]{{#1}}

\begin{document}

\selectlanguage{english}

\title{Influence of cosmological models on the GZK horizon of ultrahigh energy protons}

\author{M. De Domenico}

\address{Laboratorio sui Sistemi Complessi, Scuola Superiore di Catania, 
Via Valdisavoia 9, 95123 Catania, Italy}
\address{Istituto Nazionale di Fisica Nucleare, Sez. di Catania, 
Via S. Sofia 64, 95123 Catania, Italy}

\ead{manlio.dedomenico@ct.infn.it}

\author{A. Insolia}
\address{Istituto Nazionale di Fisica Nucleare, Sez. di Catania, 
Via S. Sofia 64, 95123 Catania, Italy}

\address{Dipartimento di Fisica e Astronomia, Universit\'a degli Studi di Catania, 
Via S. Sofia 64, 95123 Catania, Italy}


\ead{antonio.insolia@ct.infn.it}

\date{\today}

\begin{abstract}
We investigate how the density of baryonic and cold dark matter, the density of dark energy and the value of the Hubble parameter at the present time influence the propagation of ultrahigh energy protons in the nearby Universe. We take into account energy losses in the cosmic microwave radiation, the only one relevant for protons above $10^{18}$~eV, and we explore the dependence of Greisen-Zatsepin-Kuz'min (GZK) horizon on the cosmology. We investigate several cosmological scenarios, from matter dominated to energy dominated ones, and we consider the impact of uncertainties in the Hubble parameter in a $\Lambda-$Cold Dark Matter (CDM) Universe, estimated from recent observations, on the GZK horizon. The impact of the (unknown) extragalactic magnetic field on our study is discussed, as well as possible probes of the Hubble parameter attainable by current and future experiments.
\end{abstract}




\maketitle

\flushbottom

\section{Introduction}

The nucleonic component of extragalactic ultrahigh energy cosmic rays (UHECRs) above 100 EeV (1 EeV $= 10^{18}$ eV) could be subjected to a strong attenuation because of the cosmic microwave background radiation, as it was first noted by Greisen \cite{greisen1966end} and, independently, by Zatsepin and Kuz'min \cite{zatsepin1966upper}. The main consequence of such a predicted Greisen-Zatsepin-Kuzmin (GZK) effect should be to constrain the maximum propagation distance of nuclei from extragalactic sources. Moreover, if UHECRs above $\sim50$ EeV are mainly produced by extragalactic sources distributed on cosmological distance, then a flux suppression is expected at the highest energies. Recently, the Pierre Auger Collaboration \cite{abraham2008observation} and the HiRes Collaboration \cite{abbasi2008first} reported the experimental evidence of the suppression of the UHECRs spectrum with a statistical significance of about six and five standard deviations, respectively. Although such results do not provide a definitive evidence for the existence of the GZK effect, because it can be related to a change in the shape of the injection spectrum at the sources, they agree with what it is expected from the GZK effect for protons or iron nuclei. More recent observations reveal a suppression of the spectrum above 40 EeV with significance greater than 20 standard deviations \cite{abraham2010measurement}.

Within this study we investigate the impact of cosmology on the propagation of protons, and, in particular, on their GZK horizon, i.e. the distance within which 90\% of observed protons above a certain energy threshold are expected to be produced. In fact, protons with energy above 1 EeV lose energy because of photopion production due to baryonic resonances, strange particle and multipion production, and pair production by interacting with ambient photons of the cosmic microwave background, the infrared and optical backgrounds participating marginally \cite{greisen1966end,zatsepin1966upper,puget1976photonuclear,allard2005uhe,harari2006ultrahigh,hooper2007intergalactic,allard2009interactions,allard2008implications}. In the case of UHE heavier nuclei the existence of additional energy-loss processes in the microwave, infrared and optical backgrounds, due to the giant dipole resonance, the quasi-deuteron effect and the photofragmentation, drastically reduce the corresponding propagation distance, even if by an accident of nature the GZK horizon for both iron nuclei and protons is approximately the same \cite{harari2006ultrahigh}. In the following, we will only consider protons. In this case, the continuous energy-loss approximation can be safely adopted for the study of the GZK horizon \cite{harari2006ultrahigh} (and Refs. therein).

However, it is worth remarking that many sources of uncertainty in the GZK horizon of protons have been already pointed out. First, a genuine limitation to the the study of the GZK horizon signatures is provided by the relative uncertainty in the energy of UHECR events due to the experimental resolution. A second well-known (but removable) limitation is given by the continuous energy loss approximation adopted by some authors to simplify calculations. A difference of about 10\% in the estimation of the GZK horizon of protons emerges if more accurate Monte Carlo simulations are carried out instead of a simplified analytic treatment \cite{kachelriess2009gzk}. Within our study, we will show that the uncertainty in the values of cosmological parameters, as the Hubble constant at the present time, introduces a further independent uncertainty in the horizon.

Although several models for production mechanisms of UHECRs are available \cite{nagano2000observations, bhattacharjee2000origin} (and Ref. therein), \cite{hillas1984origin, hill1987ultra, berezinsky1997cosmic, berezinsky1997ultrahigh, venkatesan1997constraints, farrar1998correlation, fargion1999ultra, arons2003magnetars}, it is generally accepted that the candidate sources are extragalactic and trace the distribution of luminous matter on large scales \cite{waxman1997signature}. In particular, it has been shown that correlation with possible high redshift sources is unlikely \cite{sigl2001testing}, whereas compact sources are favored \cite{fodor2000ultrahigh, tinyakov2001correlation}: the recent result reported by the Pierre Auger Collaboration,  from observations in the southern hemisphere, experimentally supports the latter claim, showing an high correlation between the observed data and the distribution of nearby active galactic nuclei (AGN) \cite{auger2007correlation, auger2008correlation,auger2010correlation}. However, the result has not been confirmed by the HiRes Collaboration, from observations in the northern hemisphere \cite{abbasi2008search}.

Particles produced by a single source $\mathcal{S}$ with injection energy $E_{i}$ are subjected to energy losses, and they are detected on the Earth with a degraded energy $E_{f}<E_{i}$. In any study devoted to explain the current anisotropy signal or the amount of correlation with a particular catalog of sources, it is of fundamental importance to take into account a model for UHECRs injection and propagation effects. The expected number of particles from a given source at a distance $z$ and with injection spectrum $Q(E)$, can be reasonably considered to be proportional 
to its luminosity $\mathcal{L}$, to the factor $z^{-2}$ and to the attenuation factor
\begin{eqnarray}
\omega_{\text{GZK}}(z;E_{f})\propto\int_{E_{i}\(z;E_{f}\)}^{\infty}Q(E)dE,
\end{eqnarray}
accounting for energy losses. In particular, such a function, when properly normalized, estimates the probability that particles, produced \emph{at} redshift $z$ with initial energy greater or equal than $E_{i}$, might be detected with energy $E_{f}$. The surviving fraction of particles with energy above $E_{f}$ which has been produced in the nearby Universe \emph{within} a certain distance $z$, $\Omega_{\text{GZK}}(z;E_{f})$, depends on $\omega_{\text{GZK}}(z;E_{f})$ and will be defined further in the text. In this study we consider a pure power-law injection spectrum $Q(E)\propto E^{-s}$, where $s$ is the injection index, as in recent studies \cite{cuoco2006footprint,bugaev2007interactions,aublin2009discriminating,dedomenico2011bounds}.

In this work we investigate the influence of cosmology on the function $\Omega_{\text{GZK}}(z;E_{f})$. We will consider very different models of the Universe, from flat to curved ones, and different values of main cosmological parameters. In particular, we will discuss the impact of uncertainty in the Hubble parameter at the present time, according to the $\Lambda$CDM model of the Universe and to the experimental constraints obtained from recent WMAP observations \cite{wmap2011data}. However, it is worth remarking that our results will be obtained under the assumption that i) the distribution of UHECR sources is homogeneous and ii) the luminosity of such sources, emitting a proton-only composition, is known and equal for all sources.

In Sec.\,\ref{Sec-CosmologicalScenario} we briefly present the main cosmological parameters involved in our analysis and the values adopted to investigate some representative models of the Universe. In Sec.\,\ref{Sec-Propagation} we discuss in detail the energy-loss processes that UHE protons are subjected to and we show how such processes, together with cosmological models, are taken into account in the definition of $\Omega_{\text{GZK}}(z;E_{f})$. We dedicate part of this section to estimate the effect of extragalactic magnetic field on our analysis: we show how the impact of its turbulent component on the GZK horizon is small and how it can be safely neglected in this study, where only the propagation of protons is considered. In the case of heavy nuclei, the extragalactic magnetic field has a dramatic influence on their propagation, invalidating their possible use in this work as a cosmological probe. Finally, we investigate and discuss the influence of cosmological parameters on the GZK horizon, playing a fundamental role in the search of sources of UHECRs.


\section{Cosmological scenarios}\label{Sec-CosmologicalScenario}
We start by considering the Einstein equation to describe the gravitational field (including the term containing the cosmological constant $\Lambda$) in the classical General Relativity framework. Under the assumptions of an isotropic and homogeneous Universe, we consider the Friedmann-Robertson-Walker (FRW) metric with the parameter $\kappa$ accounting for the spatial curvature: $\kappa=-1$ denotes an open metric, $\kappa=0$ a flat metric and $\kappa=1$ a closed metric. Indeed, we consider the Universe as a perfect fluid: within such assumptions, the Einstein equation leads to the well known Friedmann equations. By introducing the critical density, defined as $\varrho_{c}=3H^{2}/8\pi G$, where $H$ is the time-dependent Hubble parameter and $G$ is the Newton gravitational constant, Friedmann equations can be rewritten as a function of a dimensionless density parameter, suitable for the comparison of different cosmological models.

The density parameter accounts for the matter and the energy in the Universe, and can be parametrized as the sum of different contributions. In the standard $\Lambda$CDM model, there are some contributions to $\Omega$: $\Omega_{b}$ due to baryonic matter, $\Omega_{c}$ due to cold dark matter, $\Omega_{\Lambda}$ due to dark energy, $\Omega_{r}$ due to radiation and $\Omega_{\kappa}$ for the spatial curvature. If we define the redshift $z$ by $1+z=a^{-1}(t)$, being $a(t)$ the scale factor, the first Friedmann equation can be written in terms of $z$ and of density parameters as
\begin{eqnarray}
\label{def-Hz}
\frac{H^{2}(z)}{H_{0}^{2}}=\Omega_{r}(1+z)^{4}+\Omega_{M}(1+z)^{3}+\Omega_{k}(1+z)^{2}+\Omega_{\Lambda},
\end{eqnarray}
where $\Omega_{M}=\Omega_{b}+\Omega_{c}$ is the total density of matter and $H_{0}$ is the Hubble parameter at the present time. By taking into account that the radiation density is important only in the early Universe, i.e. at high redshifts, whereas in practice it is negligible in the late Universe, the constraint $\Omega_{M}+\Omega_{\kappa}+\Omega_{\Lambda}=1$ for the density parameters can be obtained from very general considerations. Finally, changes in the expansion rate of the Universe are described by the deceleration parameter
\begin{eqnarray}
q(z)=-\frac{\ddot{a}}{aH^{2}}=\frac{H'(z)}{H(z)}(1+z)-1,
\end{eqnarray}
from which
\begin{eqnarray}
\label{def-decpar}
q(z=0)=q_{0}=\frac{1}{2}\Omega_{M}-\Omega_{\Lambda}
\end{eqnarray}
at the present time. The parameter $q_{0}$ and the density parameters described above are varied to reproduce very different cosmological models. However, it is worth remarking that we are under the assumptions of an isotropic and homogeneous Universe in the approximation of perfect fluid.


Although the most recent observations based, for instance, on WMAP measurements \cite{wmap2011data} indicate that the curvature of the Universe is very close to flat, within the present study we are also interested in investigating the impact of non-flat cosmology on the GZK horizon of UHE protons.

In a flat Universe, we consider the deceleration parameter $q_{0}$ constrained by $\Omega_{M}+\Omega_{\Lambda}=1$, whereas in a curved Universe, we consider the curvature parameter $\Omega_{\kappa}$ constrained by $\Omega_{\kappa}=1-\Omega_{M}-\Omega_{\Lambda}$. By varying two among the three parameters, we investigate different models of the Universe. In Tab.\,\ref{tab-flatuni} and Tab.\,\ref{tab-curvuni} the values of the parameters are summarized for flat and closed models, respectively, that will be adopted in the successive analysis.

\begin{table}[!h]
\centering
\begin{tabular}{lcccl}
\hline
\hline
\textbf{Name} & $\Omega_{M}$ & $\Omega_{\Lambda}$ & $q_{0}$ & \textbf{Description}\\
\hline
$\Lambda$CDM & 0.272 & 0.728 & -0.592 & Standard cosmological model \\
EdS & 1 & 0 & 0.5 & Einstein-de Sitter model \\
& & & & (matter-dominated)\\
AFU1 & 0 & 1 & -1 & Decelerating flat Universe \\
& & & & (vacuum energy-dominated)\\
AFU2 & $\frac{1}{3}$ & $\frac{2}{3}$ & -0.5 & Decelerating flat Universe \\
\hline
\hline
\end{tabular}
\caption{Models for flat Universes considered in this study and corresponding to different values of the density of matter, the density of dark energy and the deceleration parameter.}
\label{tab-flatuni}
\end{table}

\begin{table}[!h]
\centering
\begin{tabular}{lcccl}
\hline
\hline
\textbf{Name} & $\Omega_{M}$ & $\Omega_{\Lambda}$ & $\Omega_{\kappa}$ & \textbf{Description}\\
\hline
FLO1 & 0.27 & 0.93 & -0.2 & Friedmann-Lamaitre open \\
FLO2 &  & 1.23 & -0.5 &  \\
FLO3 &  & 1.43 & -0.7 &  \\
FLC1 & 0.27 & 0.53 & 0.2 & Friedmann-Lamaitre closed \\
FLC2 &  & 0.23 & 0.5 &  \\
FLC3 &  & 0.03 & 0.7 & \\
\hline
\hline
\end{tabular}
\caption{Models for curved Universes.}
\label{tab-curvuni}
\end{table}


\section{Propagation of UHE protons}\label{Sec-Propagation}

The propagation of cosmic rays is generally treated as a diffusive process described by the Ginzburg-Syrovatskii transport equation \cite{ginzburg1964origin}, and later reviews. Such equation is rather complicated, allowing an analytical solution in few cases under very restrictive assumptions. In the case of UHECRs propagating in the extragalactic space, the transport equation can be simplified by assuming time-independent diffusion coefficient and energy losses, defining a static Universe \cite{aloisio2004diffusive,aloisio2005anti,lemoine2005extragalactic,aloisio2007dip}. A more general approach, including such time dependence for ultrarelativistic particles diffusing in an expanding Universe from a single source, has been recently proposed \cite{berezinsky2007diffusion}. Other important ingredients required to correctly describe the propagation of UHECRs are cosmic background radiations and magnetic fields. As we will see in the following, background radiations are responsible for the energy loss of UHECRs, whereas magnetic fields influence their trajectory and may have a non-negligible impact on the propagation time and on the probability to reach the Earth. For the purpose of our study, we focus on the term 
\begin{eqnarray}
\frac{dE}{dt}=-\[H(t)E+b_{\text{int}}(E,t)\],
\end{eqnarray}
describing energy-loss due to the expansion of the Universe and to proton interactions as a function of time \cite{berezinsky2007diffusion}. In the following, we will consider the energy-loss rate as a function of the redshift $z$, defined by
\begin{eqnarray}
\tau^{-1}=\frac{1}{E}\frac{dE}{dz}=-\beta(z,E)\frac{dt}{dz},\nonumber
\end{eqnarray}
in a Friedmann Universe, where
\begin{eqnarray}
-\frac{dt}{dz}&=&\frac{1}{H_{0}(1+z)}\[ \Omega_{M}(1+z)^{3} + \Omega_{\Lambda} + \Omega_{\kappa}(1+z)^{2} \]^{-\frac{1}{2}}\nonumber
\end{eqnarray}
is the factor accounting for the cosmological expansion \cite{engel2001neutrinos, ave2005cosmogenic, stanev2009high}. The function $\beta(z,E)$ is related to the cooling rate of protons and it depends on the particular energy-loss process considered. In the energy interval of our interest, three main energy-loss processes can be considered: i) adiabatic, due to cosmological expansion, ii) pair production and iii) photoproduction caused by the interaction with cosmic microwave background (CMB) radiation, also known as the Greisen-Zatsepin-Kuzmin (GZK) effect \cite{greisen1966end, zatsepin1966upper}. The energy loss due to inverse Compton effect is negligible, for energies greater than $10^{17}$ eV, and it will not be considered in the following. 

We adopt the standard black body model with temperature $T_{0}\simeq 2.725$ K for the CMB. If $\epsilon$ denotes the photon energy in the observer's rest frame, the photon energy density at the present time is defined by
\begin{eqnarray}
\label{eq-cmbr-dens}
\frac{n(\epsilon)}{\epsilon^{2}}&=&\frac{\epsilon^{2}}{\pi^{2}(\hbar c)^{3}}\(\exp\[\frac{\epsilon}{k_{B}T_{0}}\]-1\)^{-1}.\nonumber
\end{eqnarray}
The adiabatic term, accounting for energy loss rate due to the expansion of the Universe, is given by
\begin{eqnarray}
\label{def-adiabloss}
\beta_{\text{rsh}}(z)&=& H_{0}\[\Omega_{M}(1+z)^{3} + \Omega_{\Lambda} + \Omega_{\kappa}(1+z)^{2} \]^{\frac{1}{2}},
\end{eqnarray}
obtained from Eq.\,(\ref{def-Hz}). Parameterizations for pair and photomeson production are discussed in the following, whereas the impact of magnetic fields on our study will be discussed at the end of this section.

\subsection{Pair production}

In the rest frame of the proton, pair production process occurs at the threshold energy $2m_{e}c^{2}\approx 1$ MeV and it plays an important role only when CMB is considered, the CIB participating marginally \cite{puget1976photonuclear}. We treat the process as a continuous energy loss, because the loss per interaction is very small, about $10^{-3}\times E$. We consider the energy loss rate $\beta_{e^{\pm}}(E)$ accounting for the pair production, because of the Bethe-Heitler interaction with ambient photons with density $n(\epsilon)$, defined \cite{blumenthal1970energy} by
\begin{eqnarray}
\beta_{e^{\pm}}(E)\propto \alpha r_{e}^{2}Z^{2}(m_{e}c^{2})^{2}\int_{2}^{\infty}d\xi n\(\frac{m_{e}c^{2}}{2\gamma}\xi\)\frac{\varphi(\xi)}{\xi^{2}}\nonumber
\end{eqnarray}
where $\gamma\approx E/m_{p}c^{2}$ is the Lorentz factor of the proton, $m_{e}$ is the electron mass, $\alpha=e^{2}/\hbar c$ is the fine-structure constant and $r_{e}=e^{2}/m_{e}c^{2}$ is the classical electron radius. In the case of a black body radiation with temperature $T_{0}$, as in the case of CMB, the energy loss obtained in the Born approximation becomes \cite{blumenthal1970energy}
\begin{eqnarray}
\beta_{e^{\pm}}(E)=\frac{\alpha r_{e}^{2}(m_{e}c^{2}k_{B}T_{0})^{2}c}{\pi^{2}\hbar^{3}c^{3}E}f(\nu),\nonumber
\end{eqnarray}
with
\begin{eqnarray}
f(\nu)=\nu^{2}\int_{2}^{\infty}d\xi \varphi(\xi) \(e^{\nu\xi}-1\)^{-1},\quad \nu=\frac{m_{e}c^{2}}{2\gamma k_{B} T_{0}}.\nonumber
\end{eqnarray}
where $k_{B}$ is the Boltzmann constant. Higher order terms of the Born approximation, proportional to $(Z\alpha v_{\pm}/c)^{m}$, where $m$ is the number of interactions with the Coulomb field, should be taken into account in the case of nuclei heavier than protons. In fact, for $Z>1$ the symmetry between produced electron and positron breaks down. Blumenthal suggested to correct the rate through the Sommerfeld factor \cite{blumenthal1970energy}, although it is only valid in the non-relavistic limit. For nuclei with $Z>1$, this correction does not agree with experimental data and a better correction is required \cite{rachen1996thesis}. By taking into account the evolution factor, after some algebra, we obtain
\begin{eqnarray}
\label{def-betapair}
\beta_{e^{\pm}}(z,E)\simeq \frac{A_{e^{\pm}}}{E^{3}}\int_{2}^{\infty}d\xi\frac{\varphi(\xi)}{\exp\[\frac{B_{e^{\pm}}}{(1+z)E}\xi\]-1},
\end{eqnarray}
where the auxiliary function $\varphi(\xi)$ is parametrized in Ref.\,\cite{chodorowski1992reaction}, masses are in units of eV/$c^{2}$ and
\begin{eqnarray}
A_{e^{\pm}}&=&\frac{\alpha^{3}m_{e}^{2}m_{p}^{2}}{4\pi^{2}\hbar}\approx 3.44\times10^{-18}\text{ EeV}^{3}\text{ s}^{-1},\nonumber\\ 
B_{e^{\pm}}&=&\frac{m_{e}m_{p}}{2k_{B}T_{0}}\approx 1.02\text{ EeV}.\nonumber
\end{eqnarray}

\subsection{Photonuclear interactions}

The probability of UHE protons to interact with CMB photons rapidly increases with proton energy. In fact, in the proton's rest frame the interaction is equivalent to a collision with a high energy photon with energy $\epsilon'$. When the energy $\epsilon'$ equals at least the pion mass $m_{\pi}c^{2}\approx 140$ MeV, the proton undergoes photomeson production and loses energy. Such a process is known as Greisen-Zatsepin-Kuzmin effect and dominates above $50-60$ EeV \cite{greisen1966end,zatsepin1966upper}. The two main channels for the interaction, close to the threshold energy, are
\begin{eqnarray}
p+\gamma \lto \Delta(1232 \text{ MeV})\lto\begin{array}{ll}
p+\pi^{0}\\
n+\pi^{+},\\
 \quad n\lto p +e^{-}+\bar{\nu}_{e}.
\end{array}\nonumber
\end{eqnarray}
involving the resonance $\Delta(1232 \text{ MeV})$. At highest energies, heavier resonances and multipion production channels are likely. The energy-loss rate due to the interaction on CMB, the dominant one above $50$ EeV, is given by
\begin{eqnarray}
\beta_{\pi}(E)&=&\frac{m_{p}^{2}}{2E^{2}}\int_{0}^{\infty}d\epsilon\frac{n(\epsilon)}{\epsilon^{2}}\int_{0}^{2\epsilon\frac{E}{m_{p}}}d\epsilon' \epsilon'\mathcal{K}(\epsilon')\sigma(\epsilon')\nonumber\\
&=&-\frac{k_{B}T_{0}}{2\pi^{2}\hbar}\frac{m^{2}_{p}}{E^{2}}\int_{0}^{\infty}d\epsilon \mathcal{K}(\epsilon)\sigma(\epsilon)\epsilon\times\ln\[1-\exp\(-\frac{m_{p}}{2Ek_{B}T_{0}}\epsilon\)\],\nonumber
\end{eqnarray}
where $m_{p}$ is the proton mass in units of eV$/c^{2}$, $\sigma(\epsilon)$ is the cross-section for pion production in terms of the photon energy $\epsilon$ and $\mathcal{K}(\epsilon)$ is the inelasticity factor \cite{stecker1968effect,rachen1993extragalactic}. 

\begin{figure}[!t]
	\centering
	  \includegraphics[width=11.0cm]{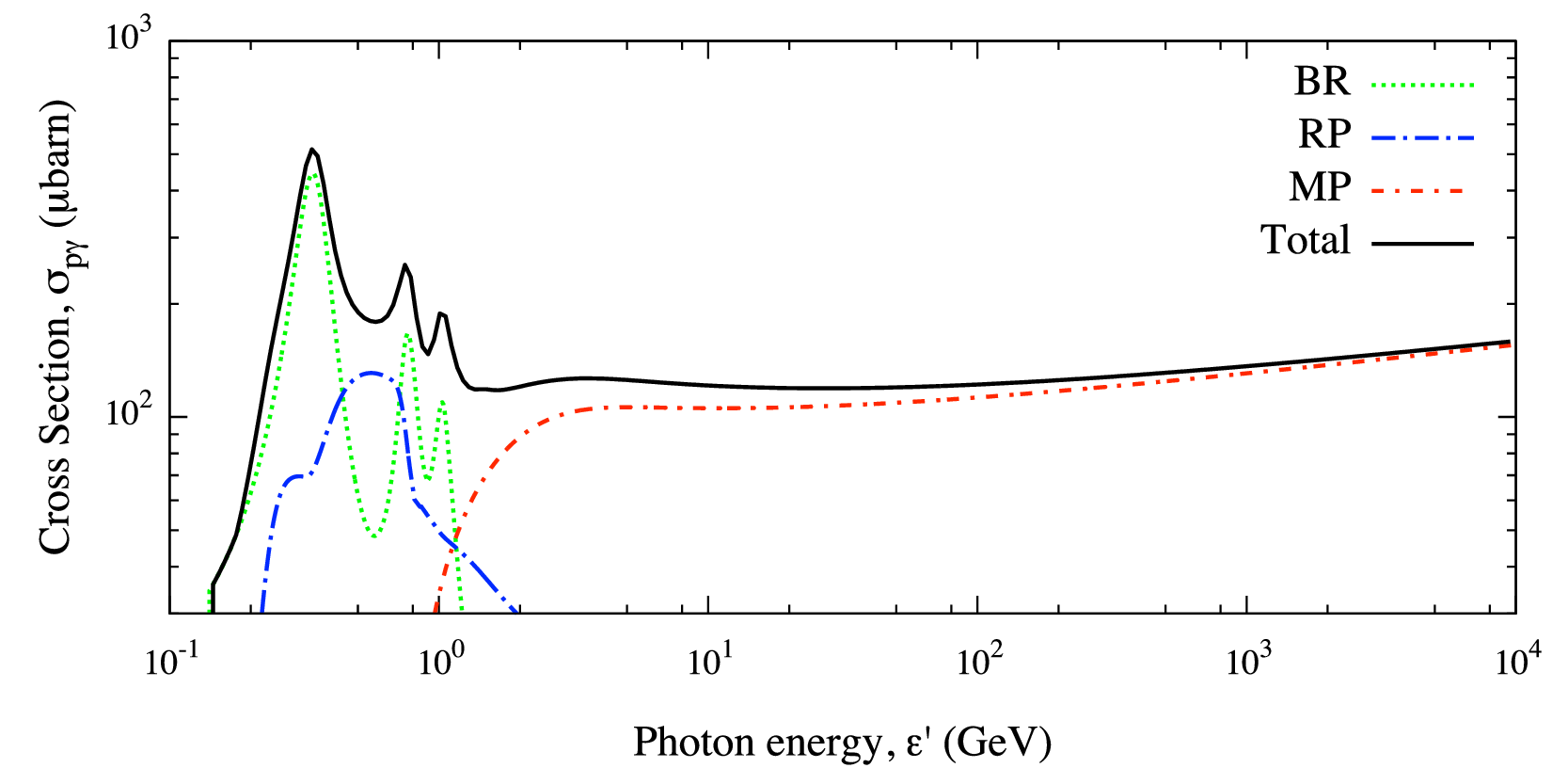}
	\caption{Cross-section for the photomeson production of a proton in the CMB radiation. Contributions from baryonic resonances (BR), Rachen's parameterizations for direct particle production (RP) and multipions (MP) are shown. Energy has to be considered in the rest frame of the proton.}
\label{fig-crossec}
\end{figure}

Just above the threshold, baryonic resonances dominate and protons are subjected to photomeson production, mainly through the $\Delta(1232)$-baryon resonance, whereas heavier resonances (up to $\Delta(1950)$-baryon) play a more marginal role. The cross-section for baryonic resonances is parametrized by
\begin{eqnarray}
\sigma_{\text{BR}}(\epsilon)&=& \sum_{i=1}^{4}\sigma_{i}\sigma_{L}(\epsilon;\epsilon_{i},\Gamma_{i})\nonumber
\end{eqnarray}
where $\sigma_{L}$ is the Lorentzian function, ($\epsilon_{i}$ (GeV), $\Gamma_{i}$ (GeV), $\sigma_{i}$ ($\mu$b)) $=(0.34, 0.17, 351)$, $(0.75, 0.50,159)$, $(1.00, 0.60, 21)$ and $(1.50, 0.80, 26)$ for $i=1,2,3$ and 4, respectively. For all other processes participating in photomeson production, including multipions (MP) or direct particle production involving $\pi$, $\eta$, $\Delta$, $\rho$, $\omega$ and strange-particle channels (RP), we use Rachen's parameterizations \cite{rachen1996thesis}. In Fig.\,\ref{fig-crossec} are shown the cross-sections for the discussed $p\gamma$ interactions, separately, and the total cross-section. 

By taking into account the evolution factor, the energy-loss rate for photomeson production is given by $\beta_{\pi}(z,E)=(1+z)^{3}\beta_{\pi}(z=0,E(1+z))$ for CMB. Finally, the total energy-loss rate is defined as
\begin{eqnarray}
\label{def-energylosseq}
\frac{1}{E}\frac{dE}{dz} = -\frac{dt}{dz}\times\[ \beta_{\text{rsh}}(z) + \beta_{\pi}(z,E) + \beta_{e^{\pm}}(z,E) \],
\end{eqnarray}
taking into account the contributions of the processes discussed in this section. Efficient analytical parameterizations of the cooling rates can be considered for the numerical estimation of the energy loss. First, we have verified that the recent parameterization \cite{cuoco2006footprint}
\begin{eqnarray}
\beta_{\pi}(z,E)\simeq\left\{
\begin{array}{ll}
A_{\pi}(1+z)^{3}\exp\[\frac{B_{\pi}}{(1+z)E}\] & E\leq \tilde{E}(z)\\
C_{\pi}(1+z)^{3} & E> \tilde{E}(z)
\end{array}\right.
\end{eqnarray}
provides an excellent approximation for the energy-loss rate due to photomeson production. Here, the function $\tilde{E}(z)=6.86 e^{-0.807z}\times 10^{20}$ eV ensures the continuity of $\beta_{\pi}(z,E)$ and $\{A_{\pi},B_{\pi},C_{\pi}\}=\{3.66\times10^{-8}\text{yr}^{-1},2.87\times10^{20}\text{eV},2.42\times10^{-8}\text{yr}^{-1}\}$ are taken from Ref. \cite{anchordoqui1997effect}. 

\begin{figure}[!t]
	\centering
	  \includegraphics[width=11.0cm]{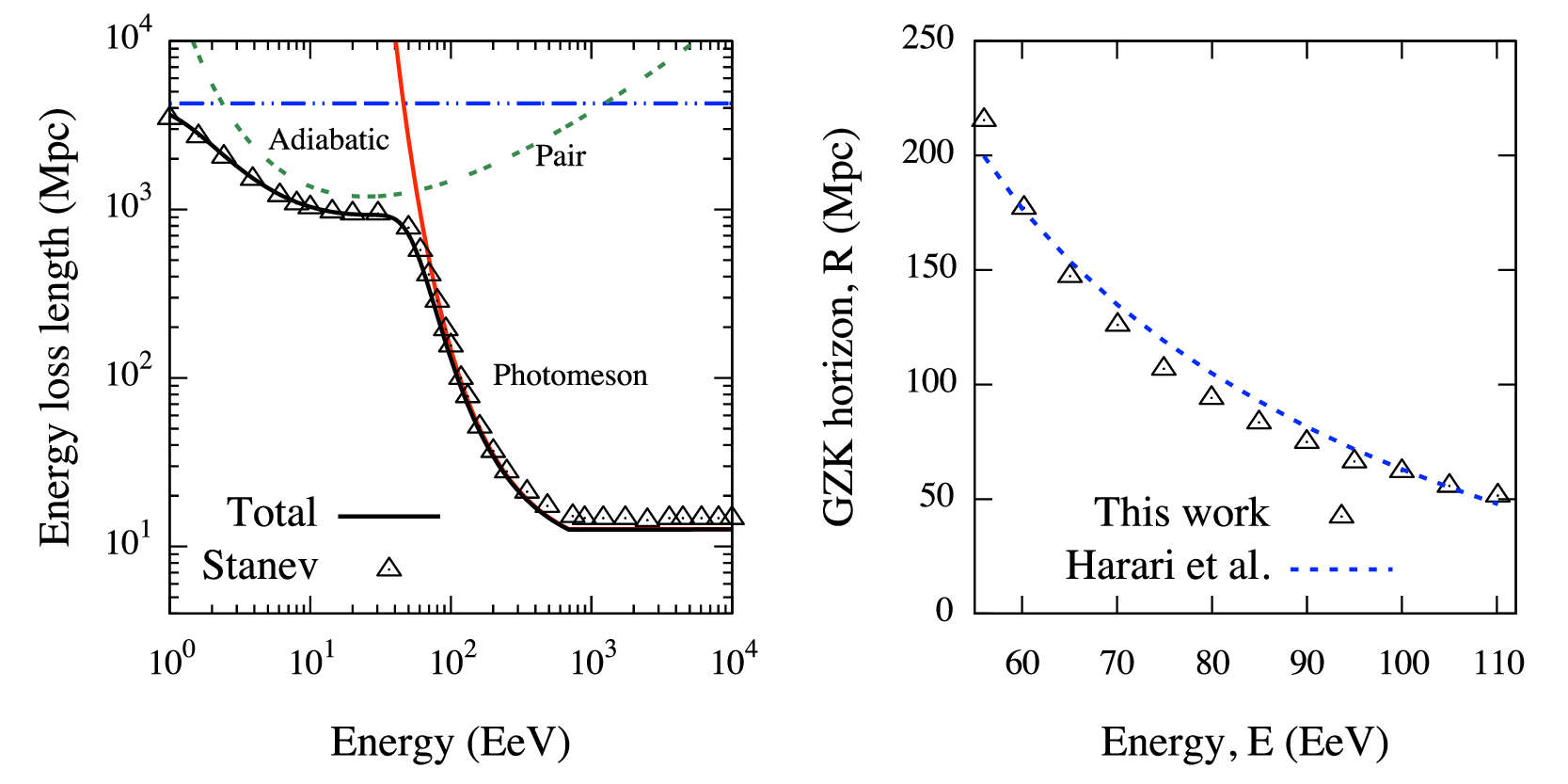}
	\caption{\emph{Left panel:} Energy loss length in the CMB, for each process separately and for all processes together, in the case of a proton; triangles indicate the values recently reported by Stanev \cite{stanev2009propagation}. \emph{Right panel:} GZK horizon as a function of the propagated proton energy (triangles) and values obtained by Harari \emph{et al} \cite{harari2006ultrahigh} (dashed line).}
\label{fig-enloss}
\end{figure}

In Fig.\,\ref{fig-enloss} (left panel) is shown the energy-loss length in the CMB, for each process separately and for all processes together, in the case of a proton propagating in a $\Lambda$CDM Universe. In the energy interval between 1 EeV and $50-60$ EeV the main energy-loss process is the pair production, whereas photomeson production dominates up to the highest energies. We show, for comparison, the excellent agreement with the total loss length recently reported by Stanev, for a proton propagating in the CMB \cite{stanev2009propagation}. 

The solution of Eq.\,(\ref{def-energylosseq}) is the degraded energy $E_{f}$ of the proton with initial energy $E_{i}$, after the propagation from the source at redshift $z$ to the Earth ($z=0$). Under the assumption of a power-law injection spectrum, we have the surviving probability
\begin{eqnarray}
\label{def-wgzk}
\omega_{\text{GZK}}(z;E_{f})=\frac{s-1}{E_{f}^{-s+1}}\int_{E_{i}\(z;E_{f}\)}^{\infty}E^{-s}dE,
\end{eqnarray}
while assuming equal intrinsic luminosity and homogenous distribution of sources (i.e. their number at redshift $z$ is proportional to $z^{2}$), we obtain the surviving flux defined by
\begin{eqnarray}
\label{def-wgzk-flux}
\Omega_{\text{GZK}}(z;E_{f})=\frac{\int_{z}^{\infty}dz'\int_{E_{i}\(z';E_{f}\)}^{\infty}E^{-s}dE}{\int_{0}^{\infty}dz'\int_{E_{i}\(z';E_{f}\)}^{\infty}E^{-s}dE},
\end{eqnarray}
for the probability of detecting at $z=0$ a proton emitted with $E\geq E_{f}$ farther than a distance $z$. The energy $E_{i}\(z;E_{f}\)$ of the injected proton is estimated by evolving Eq.\,(\ref{def-energylosseq}) backward in time \cite{harari2006ultrahigh,aublin2009discriminating}. The GZK horizon $R$ is the distance such that $1-\Omega_{\text{GZK}}(R,E_{f})=0.9$. In Fig.\,\ref{fig-enloss} (right panel) is shown the GZK horizon as a function of the propagated proton energy and we show, for comparison, the good agreement with results obtained by Harari \emph{et al} \cite{harari2006ultrahigh}.

\subsection{Extragalactic magnetic field}

It is worthwhile discussing in this section the impact of the extragalactic magnetic field (EMF) on the probability $\Omega_{\text{GZK}}(z;E_{f})$. Unfortunately, a direct or indirect measurement of the EMF is still missing, and only bounds to its r.m.s. strength $B_{\text{rms}}$ have been estimated. However, bounds on the EMF strength also depend on the field correlation length $\ell$, which is also unknown.

Upper limits of the order of $10^{-9}$~G (1~nG) on the intensity of the EMF have been measured through i) the Faraday rotation in the polarized radio emission from distant quasars \cite{vallee2004cosmic} and ii) the characteristic distortions that it induced on the spectrum and the polarization of CMB radiation \cite{seshadri2009cosmic}. More recent analyses, based on the recent WMAP measurements \cite{wmap2011data}, provide the more stringent upper bound of $\approx$2~nG on the present value of the cosmic magnetic field of primordial origin, more than one order of magnitude smaller than previous estimates \cite{trivedi2010primordial}.

Intergalactic magnetic field might also be structured inside and around clusters or groups of galaxies, with filaments extending over few Mpc, as shown, for instance, in recent detailed simulations \cite{ryu2008turbulence}. The topology of such a structured magnetic field would have a non-negligible impact on the trajectories of UHE protons and, of course, on the average deflections they experience in the case of EMF with regular structures above 200~kpc \cite{das2008propagation}. Additionally, longitude-averaged X-ray emission observed with ROSAT near 0.65~keV and 0.85~keV towards the center of the Galaxy, are in agreement with a Galactic wind thermally-driven by cosmic rays and hot gas \cite{everett2008milky,everett2010synchrotron}. In our study, we can neglect the effect of such a magnetic wind, because it is expected to have a non-negligible impact for protons below 60~EeV or heavier nuclei. Moreover, we neglect the case of a structured EMF because of the lack of direct or indirect measurements about its structure. 

\begin{figure}[!t]
	\centering
	  \includegraphics[width=11.0cm]{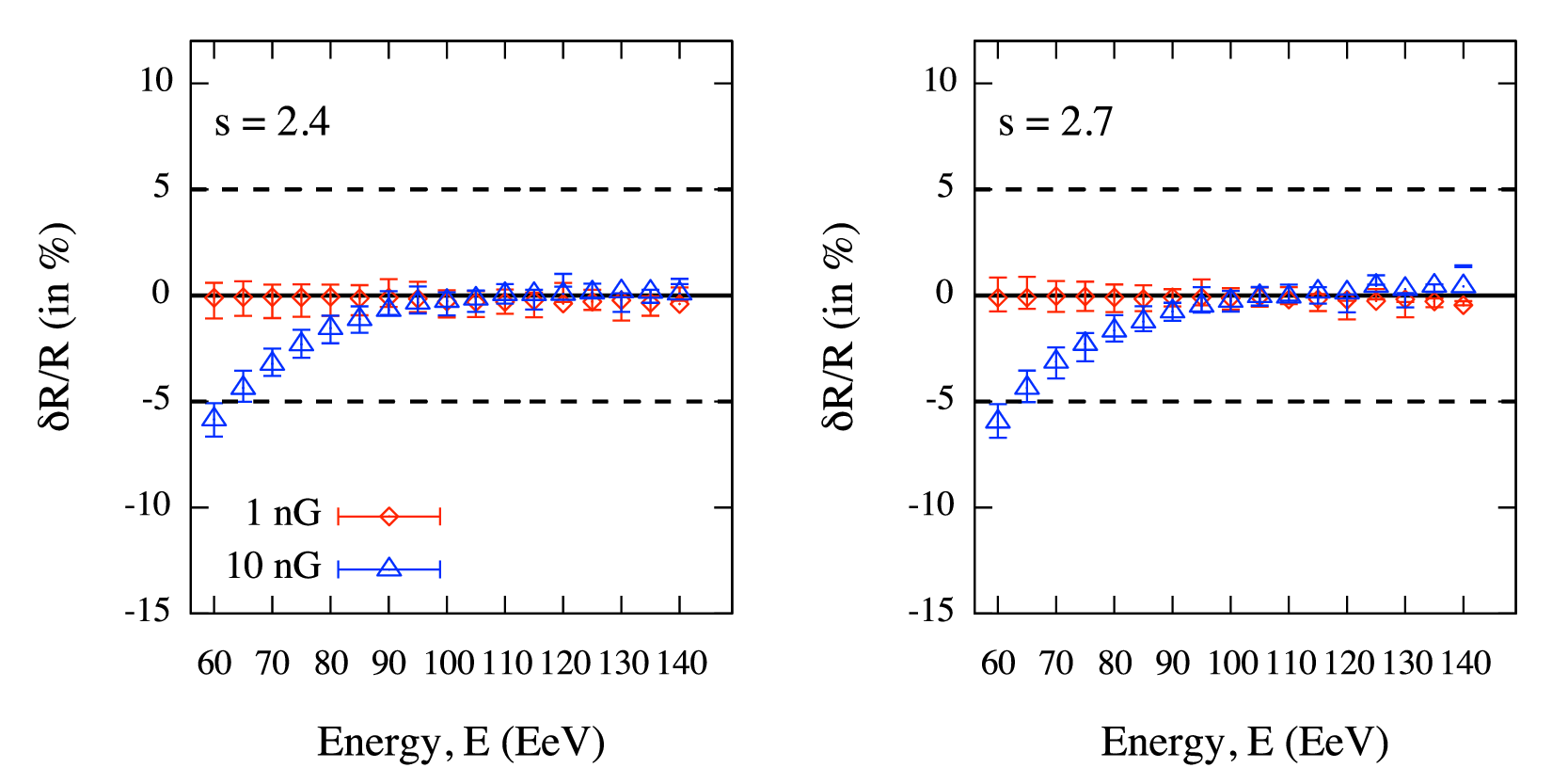}
	\caption{Relative difference between GZK horizons estimated with and without EMF. Two values of the r.m.s. strength of the EMF are considered, namely 1 and 10~nG, as well as two different injection index, namely 2.4 (left panel) and 2.7 (right panel). Error bars indicates only the statistical uncertainty on $\delta R/R$. Lines corresponding to $\pm5\%$ bands and  $\delta R/R=0$ are shown for reference.}
\label{fig-gzk-rel-diff}
\end{figure}

Here, we consider only the case of a turbulent EMF characterized by $B_{\text{rms}}$ and $\ell$. In this case, protons propagate randomly following a brownian trajectory. The average deflection they are subjected to can be parameterized by
\begin{eqnarray}
\label{def-sigl-theta}
\theta(E,D)\simeq 0.8^{\circ}\(\frac{E}{10^{20}~\text{eV}}\)^{-1}\(\frac{D}{10~\text{Mpc}}\)^{\frac{1}{2}}\(\frac{\ell}{1~\text{Mpc}}\)^{\frac{1}{2}}\(\frac{B_{\text{rms}}}{10^{-9}~\text{G}}\),
\end{eqnarray}
in absence of energy losses \cite{bhattacharjee2000origin}, with $D\approx zc/H_{0}$ for distances considered in the present study. Such a deflection implies an average time delay
\begin{eqnarray}
\label{def-sigl-tau}
\tau(E,D)\simeq 1.5\(\frac{E}{10^{20}~\text{eV}}\)^{-2}\(\frac{D}{10~\text{Mpc}}\)^{2}\(\frac{\ell}{1~\text{Mpc}}\)\(\frac{B_{\text{rms}}}{10^{-9}~\text{G}}\)^{2}~\text{kyr},
\end{eqnarray}
relative to the rectilinear propagation with the speed of light. Let us consider a propagation distance equal to the GZK radius, i.e. $D=R$: the difference between the trajectory corresponding to the propagation in absence of magnetic field and the brownian trajectory of the proton propagating in the EMF is $\delta R \approx c\tau$, leading to
\begin{eqnarray}
\label{def-deltaR-R}
\frac{\delta R}{R}\simeq 0.5\times 10^{-4}\(\frac{E}{10^{20}~\text{eV}}\)^{-2}\(\frac{R}{10~\text{Mpc}}\)\(\frac{\ell}{1~\text{Mpc}}\)\(\frac{B_{\text{rms}}}{10^{-9}~\text{G}}\)^{2},
\end{eqnarray}
where we have omitted to explicitly report the dependence of $R$ and $\delta R$ on the energy $E$, for simplicity. We estimate the bound to $B_{\text{rms}}\sqrt{\ell}$ in order to have a negligible impact on the GZK horizon, i.e. $\delta R/R<5\%$, by
\begin{eqnarray}
\sqrt{\frac{\ell}{1~\text{Mpc}}}\frac{B_{\text{rms}}}{10^{-9}~\text{G}}<10^{\frac{3}{2}} \(\frac{E}{10^{20}~\text{eV}}\)\(\frac{R}{10~\text{Mpc}}\)^{-\frac{1}{2}}.
\end{eqnarray}
For a UHE proton with $E=60$~EeV the GZK horizon is $R\approx180$~Mpc (for $H_{0}=70.4$~km/s$^{-1}$/Mpc): if we consider the propagation in an EMF characterized by $\ell=1$~Mpc, any r.m.s. strength $B_{\text{rms}}<4.5$~nG will not significantly affect the horizon. For a proton with higher energy as 95~EeV, such a bound reads $B_{\text{rms}}<11.5$~nG, i.e. even more intense EMFs have a negligible impact on the result. For instance, the estimated upper bounds are still conservative even if larger correlation length as 16~Mpc are considered, reducing to $\approx$1~nG and $\approx$3~nG, respectively.

We have investigated the validity of such an argument by performing several Monte Carlo realizations of protons propagating in the Universe, with and without EMF. For such a purpose, the well known propagation software CRPropa \cite{armengaud2007crpropa} has been adopted: two different values of the r.m.s. strength of the EMF, namely 1 and 10~nG, as well as two different values of the injection index, namely 2.4 and 2.7, has been considered for this study. In any case, the correlation length has been fixed to $\ell=1$~Mpc. The main advantage in using CRPropa is that relevant energy-loss processes are considered during the propagation in a turbulent magnetic field: in this case we can quantify the average relative deviation $\delta R/R$ in a more realistic scenario with respect to Eq.\,(\ref{def-deltaR-R}), where energy loss has been switched off to estimate $\theta(E,D)$ and $\tau(E,D)$. In Fig.\,\ref{fig-gzk-rel-diff} is shown the result of such a study. It is evident that the the injection index has a negligible impact on the relative deviation $\delta R/R$. For both $B_{\text{rms}}\simeq1$~nG and $B_{\text{rms}}\simeq10$~nG, the resulting relative deviation is within 5\% at any energy, in agreement with expectation. It is worthwhile noticing that in any case, the value of $\delta R/R$ is non-positive, i.e. the GZK horizon for the propagation in an EMF is smaller than the horizon for the propagation without magnetic field, as axpected. Moreover, it is relevant to remark that for the weak EMF (1~nG) $\delta R/R\approx 0$, whereas for the strong EMF (10~nG) $\delta R/R$ approaches zero above 90~EeV: the EMF has no impact on the GZK horizon of UHE protons.

From such a result, we can also deduce that above 60~EeV the effect of energy loss on the propagation in an EMF is negligible: hence, the parameterizations given by Eq.\,(\ref{def-sigl-theta}) and (\ref{def-sigl-tau}), although they have been obtained by considering no energy losses, can be safely used in the scenarios previously described.

Such a result is rather conservative: in fact, the most up-to-date bounds on the correlation length and the r.m.s. strength of the EMF are much lower of the values considered in this study \cite{neronov2010evidence}.


\section{Analysis and discussion}

\begin{figure}[!b]
	\centering
	  \includegraphics[width=11.0cm]{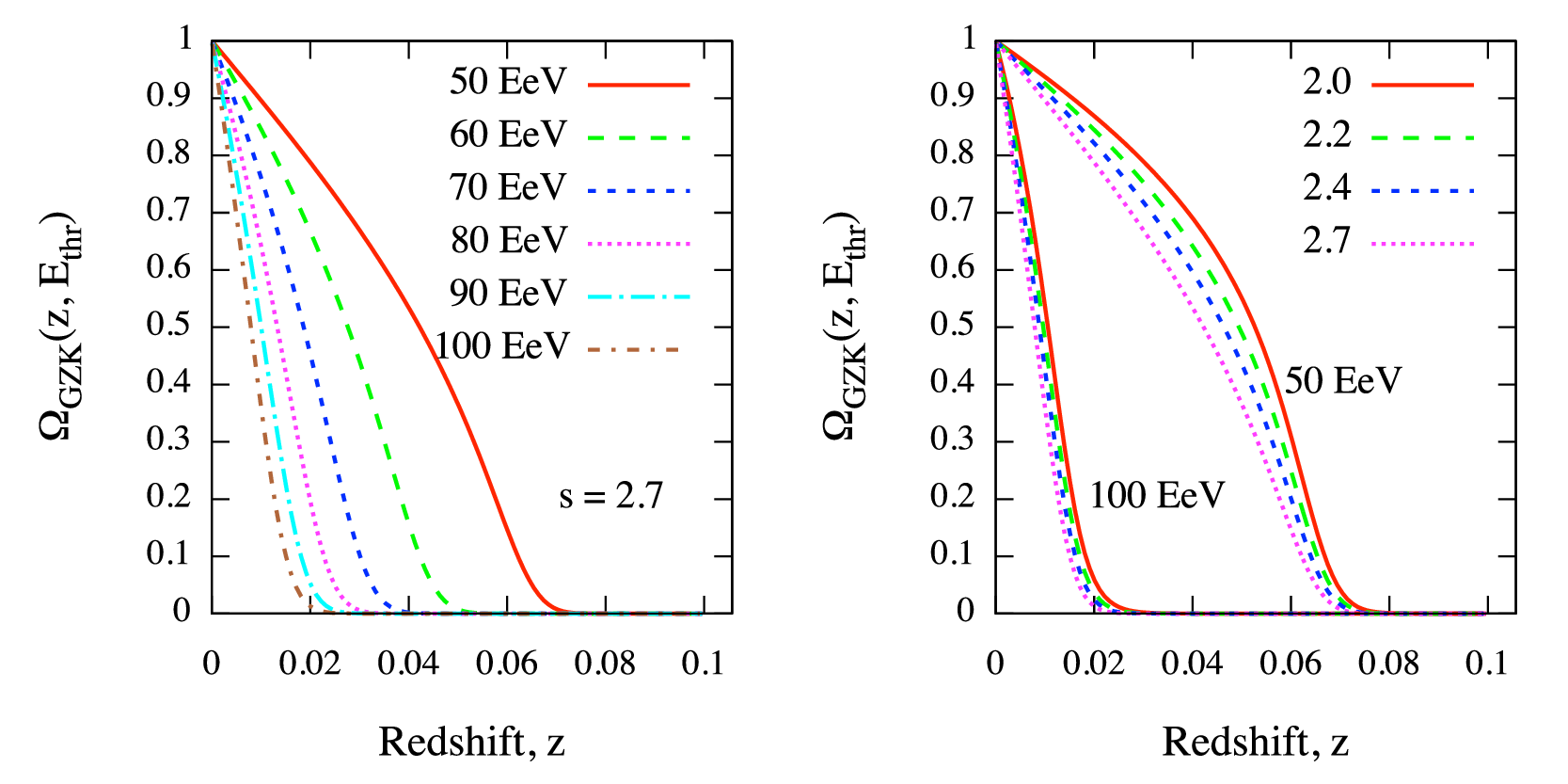}
	\caption{Surviving flux, defined by Eq.\,(\ref{def-wgzk-flux}), due to cosmological effects described in the text, in the case of a proton and assuming $\Lambda$CDM model parameters. \emph{Left panel:} the value of the injection index is fixed ($s=2.7$) while the energy threshold $E_{f}$ is varied; \emph{Right panel:} the value of the energy threshold $E_{f}$ is fixed (two cases, $E=50$ EeV and $E=100$ EeV) while the injection index is varied.}
\label{fig-wgzkES}
\end{figure}

In Fig.\,\ref{fig-wgzkES} is shown the function $\Omega_{\text{GZK}}(z;E_{f})$ for some values of the energy threshold $E_{f}=E_{\text{thr}}$, as a function of the distance. By assuming the most recent values of $\Lambda$CDM model parameters ($\Omega_{b}=0.0456$, $\Omega_{c}=0.227$, $\Omega_{\Lambda}=0.728$ and $H_{0}=70.4$ km/s$^{-1}$/Mpc) \cite{wmap2011data}, the value of the injection index is fixed ($s=2.7$) while the energy threshold $E_{\text{thr}}$ is varied (left panel): as expected, the GZK horizon decreases by increasing the energy threshold. In the right panel of Fig.\,\ref{fig-wgzkES}, is shown the same function for two fixed values of the energy threshold, $E_{\text{thr}}=50$ EeV and $E_{\text{thr}}=100$ EeV, while the injection index is varied: although differences among curves corresponding to different values of the injection index are not exaggerated as in the previous case, we find that they increase by decreasing the energy threshold $E_{\text{thr}}$. Although observations suggest an injection index between 2.2 and 2.6, depending on the underlying assumptions, we have extended our study to a broader range, namely from 2.0 to 2.7, because of interest for recent studies \cite{dedomenico2011bounds,ahlers2011cosmogenic,decerprit2011constraints}.

Cosmological parameters as the density of matter, the density of dark energy and the Hubble parameter at the present time, should influence the function $E_{i}\(z;E_{f}\)$, discussed at the end of the previous section, and by consequence the probability $\Omega_{\text{GZK}}(z;E_{f})$. From Eq.\,(\ref{def-energylosseq}) it is clear that the factor involving such parameters is only $-dt/dz$. 

For small values of $z$, corresponding to the nearby Universe, it is not expected a significant difference among cosmological models. At the highest energies, above $50$ EeV, we have found (see Fig.\,\ref{fig-enloss}, left panel) that photomeson production dominates energy losses, thus only the term $-\beta_{\pi}(z,E)\times dt/dz$ is important. By considering two extreme scenarios, as matter-dominated ($\Omega_{M}=1; \Omega_{\Lambda}=0$) or energy-dominated ($\Omega_{M}=0; \Omega_{\Lambda}=1$) Universes, the energy-loss term is modified at most by 10\%. 

\begin{figure}[!htb]
	\centering
	  \includegraphics[width=11.0cm]{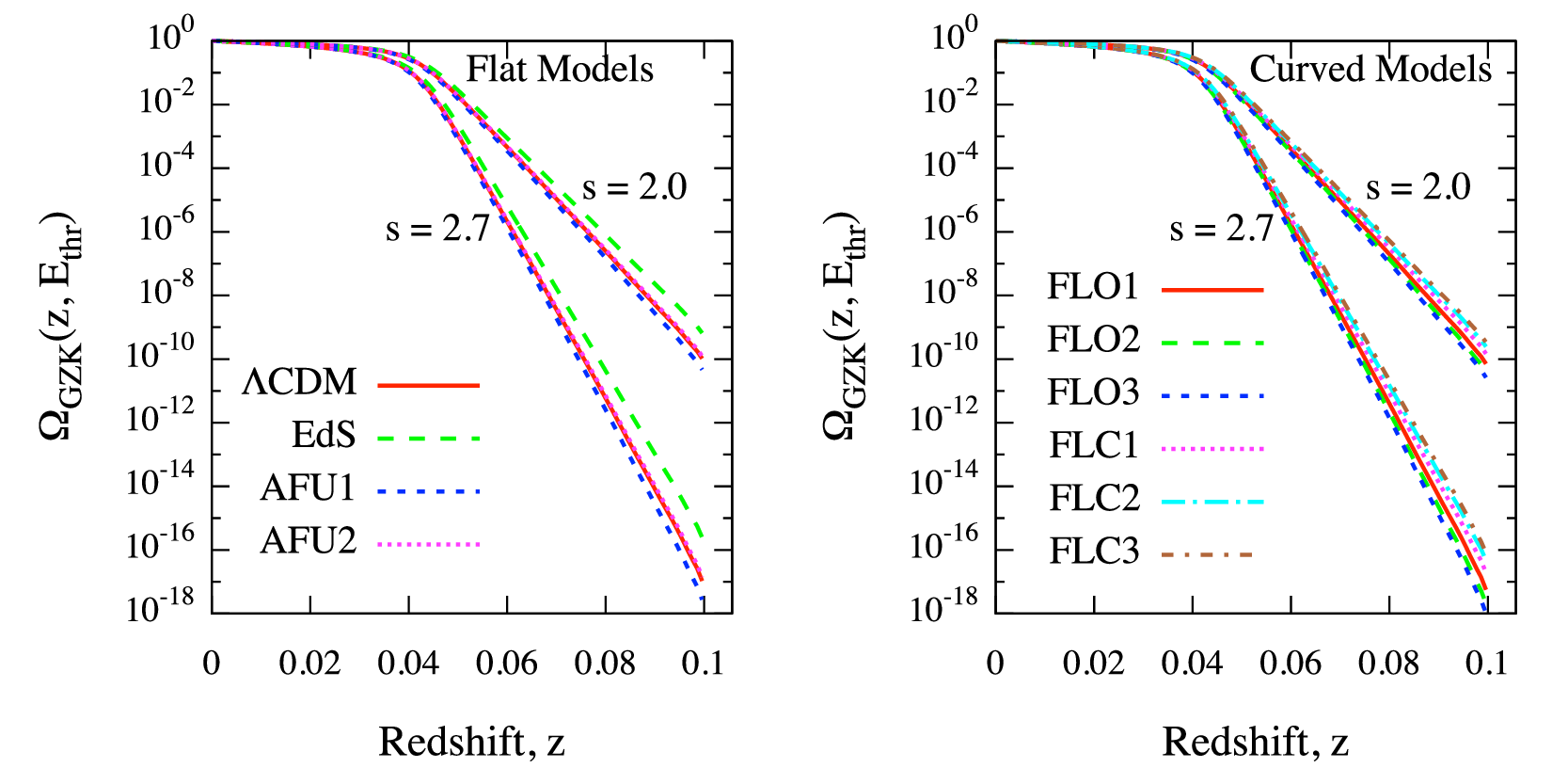}
	\caption{Surviving flux, defined by Eq.\,(\ref{def-wgzk-flux}), due to cosmological effects described in the text, in the case of a proton with $E_{\text{thr}}=60$~EeV and assuming different models of the Universe, for two values of the injection index ($s=2.0$ and $s=2.7$). \emph{Left panel:} flat models described in Tab.\,\ref{tab-flatuni} are considered; \emph{Right panel:} curved models described in Tab.\,\ref{tab-curvuni} are considered.}
\label{fig-wgzkUniverseModels}
\end{figure}

In Fig.\,\ref{fig-wgzkUniverseModels} is shown the function $\Omega_{\text{GZK}}(z;E_{f})$ as a function of the redshift, for both flat and curved models listed in Tab.\,\ref{tab-flatuni} and \ref{tab-curvuni}, respectively. With no regards for the value of the injection index $s$, functions corresponding to very different cosmological models do not differ significantly below the GZK horizon, which is not influenced by the choice of a particular cosmology. The difference becomes significant toward higher redshifts where, however, the surviving probability for a proton is very small.

\begin{figure}[!t]
	\centering
	  \includegraphics[width=11.0cm]{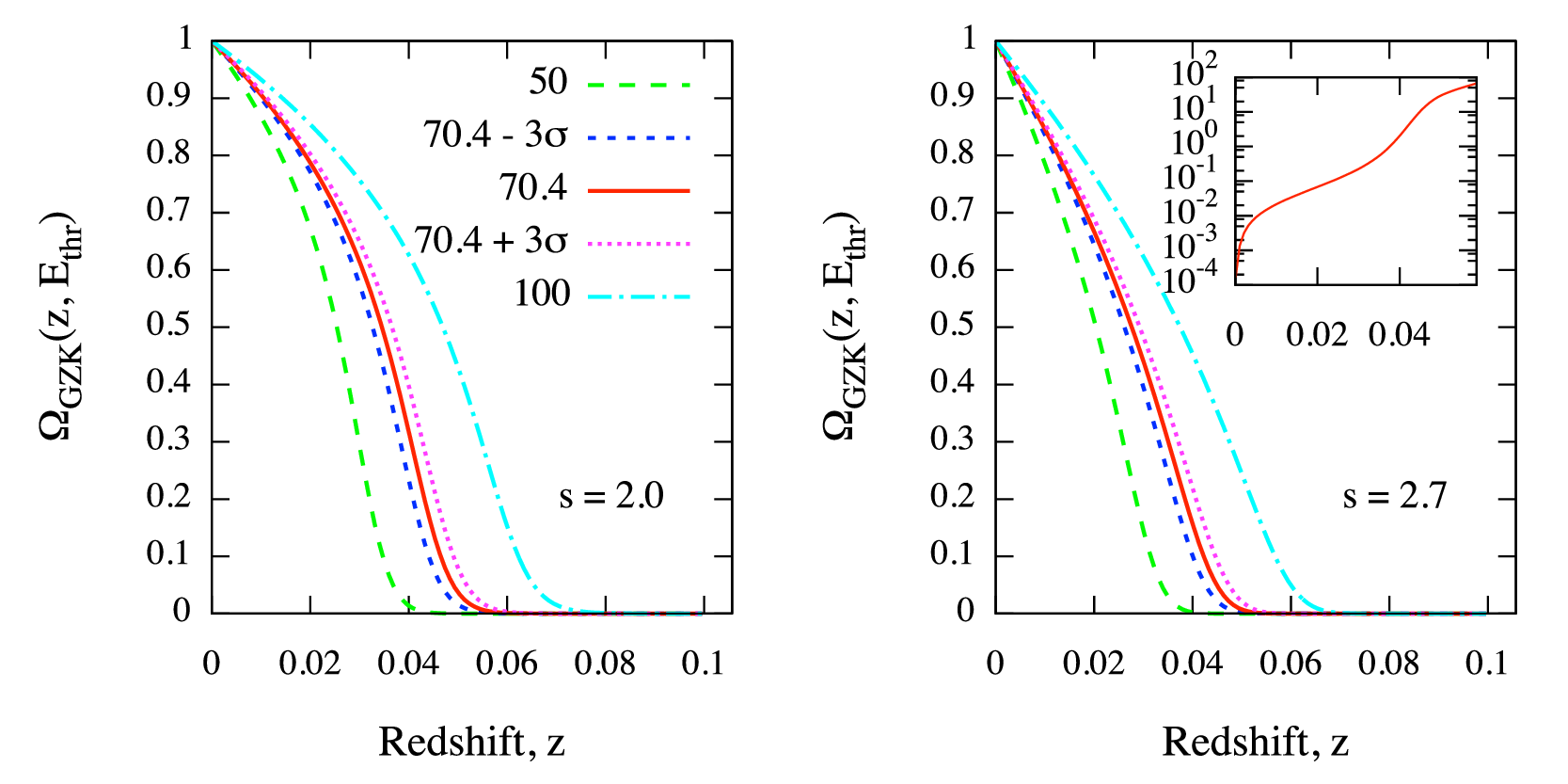}
	\caption{Surviving flux, as in Fig.\,\ref{fig-wgzkUniverseModels}, assuming $\Lambda$CDM model values for density parameters and protons with energy threshold $E_{\text{thr}}=60$ EeV, for two values of the injection index ($s=2.0$ and $s=2.7$). Different values of the Hubble parameter $H_{0}$ (in km/s$^{-1}$/Mpc units) at the present time are considered. The inset in the right panel shows the relative difference between curves corresponding to $H_{0}^{+}=70.4+3\sigma\approx75$ and $H_{0}^{-}=70.4-3\sigma\approx66$, as a function of the distance.}
\label{fig-wgzkH}
\end{figure}

Indeed, in Fig.\,\ref{fig-wgzkH} is shown the same function by assuming $\Lambda$CDM model values for density parameters \cite{wmap2011data}, protons with energy threshold $E_{\text{thr}}=60$ EeV and different values of the Hubble constant $H_{0}$ at the present time. For simplicity, in the following we will not explicitly specify the unit of $H_{0}$, that should be considered km/s$^{-1}$/Mpc. Left panel shows the case for $s=2.0$, whereas right panel shows the case for $s=2.7$. It is evident that the value of the Hubble parameter significantly affects the function $\Omega_{\text{GZK}}(z;E_{f})$, as already suggested by previous studies \cite{rachen1993extragalactic,cuoco2006footprint}. If the $3\sigma$ uncertainty in the value ($H_{0}=70.4$) obtained from the $\Lambda$CDM model is taken into account \cite{wmap2011data}, as shown in both panels of Fig.\,\ref{fig-wgzkH}, we find that differences between curves corresponding to $70.4+3\sigma$ and $70.4-3\sigma$ are negligible ($\approx 5\%$) only for $z<0.018$, as shown in the inset of the right panel, for $s=2.7$. From the lower bound curve ($H_{0}^{-}=70.4-3\sigma\approx66$) the estimated GZK horizon is $z\approx0.039$ ($\approx 156$~Mpc), whereas for the upper bound curve ($H_{0}^{+}=70.4+3\sigma\approx75$) it is $z\approx0.044$ ($\approx198$~Mpc), with a relative difference of about 11\% in redshift and about 21\% in distance. A relative difference in redshift of about 100\% between the corresponding surviving functions is reached around $z=0.037$, tending to increase with the redshift.

\begin{figure}[!t]
	\centering
	  \includegraphics[width=11.0cm]{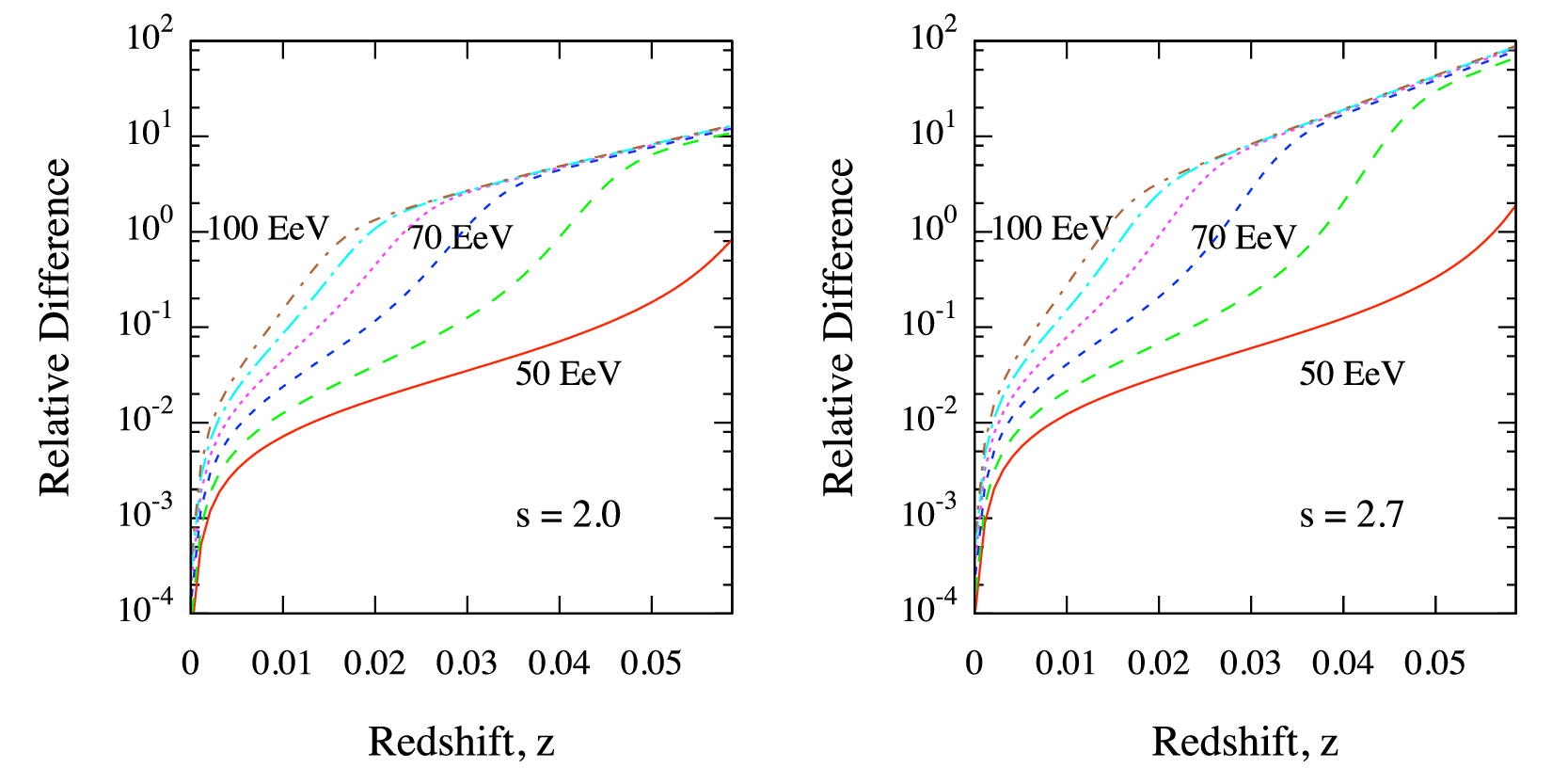}
	\caption{Relative difference between curves corresponding to probability functions for $H_{0}^{+}$ and $H_{0}^{-}$ in Fig.\,\ref{fig-wgzkH}, for different values of energy threshold in the case of $s=2.0$ (left panel) and $s=2.7$ (right panel).}
\label{fig-reldiff}
\end{figure}

In Fig.\,\ref{fig-reldiff} are shown the relative differences between curves corresponding to surviving functions for $H_{0}^{+}$ and $H_{0}^{-}$, by varying the energy threshold and the injection index, $s=2.0$ (left panel) and $s=2.7$ (right panel). By increasing the energy threshold the relative difference increases, although it keeps smaller than 15\% below the GZK horizon. All significant differences are found to be above the GZK horizon, beyond which only 10\% of protons output by their sources are able to reach the Earth. In the following we will discuss two simple applications of the arguments discussed so far, in order to show how current and future experiments could probe the value of the Hubble parameter.

\subsection{Application \#1: energy spectrum}

\begin{figure}[!t]
	\centering
	  \includegraphics[width=11.0cm]{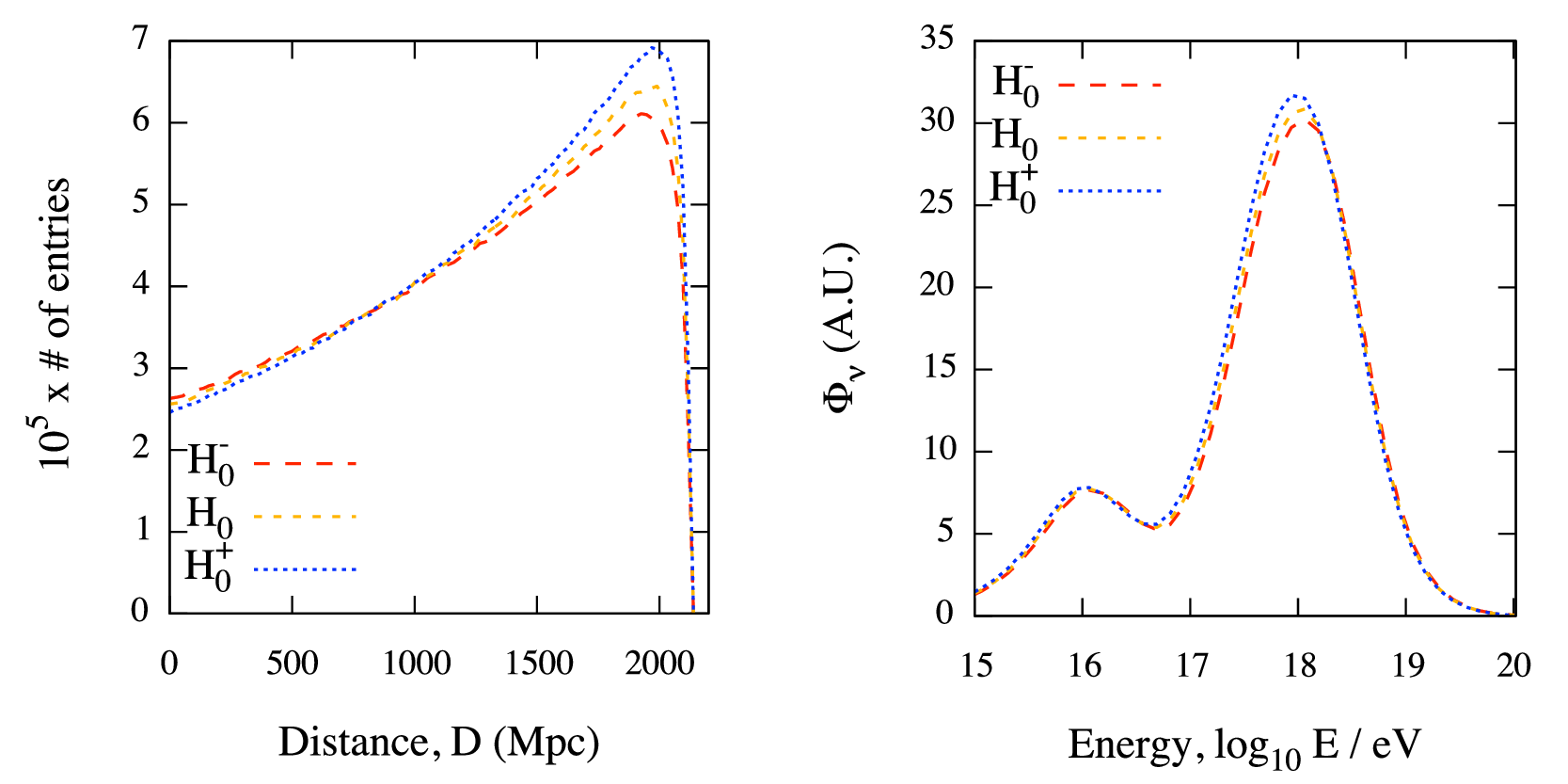}
	\caption{UHE protons produced by sources homogeneously distributed within 2~Gpc (see the text for further detail); injection spectrum follows a power-law with spectral index $s=2.7$ and $E_{\text{max}}=1000$~EeV. \emph{Left panel}: number of entries versus the distance at which GZK neutrinos are generated, by varying the Hubble parameter at present time; $H_{0}=70.4$, $H_{0}^{-}$ and $H_{0}^{+}$ are considered. \emph{Right panel:} corresponding fluxes of GZK neutrinos at Earth.}
\label{fig-gzk-nu}
\end{figure}

As a first practical application, we consider the propagation of $10^{8}$ UHE protons produced with CRPropa \cite{armengaud2007crpropa} by sources homogeneously distributed within 2~Gpc. It is worth remarking that CRPropa performs propagation of particles with respect to distance, for a specified astrophysical and cosmological scenario. Hence, in this section, we will report the results with respect to distance. Moreover, we consider neutrinos produced by UHE protons and propagated by CRPropa. The injection spectrum at source is considered to be $\propto E^{-2.7}$, with a maximum energy of $E_{\text{max}}=1000$~EeV. The propagation of protons that have reached an energy smaller than 10~EeV, because of the interactions with CMB photons, is not followed. In particular, we consider three astrophysical scenarios, with different values of the Hubble parameter at the present time, namely $H_{0}=70.4$, $H_{0}^{-}$ and $H_{0}^{+}$. In Fig.\,\ref{fig-gzk-nu} (left panel) is shown the number of entries versus the distance at which GZK neutrinos are generated, together with the flux of GZK neutrinos at Earth, for each astrophysical scenario (right panel). While for distances below $\approx1$~Gpc, differences in the three scenarios are small, at large distances, comparable to those of the most far sources, a larger number of neutrinos is created for increasing values of the Hubble parameter. 

The interaction rate is expected to decrease for increasing values of $H_{0}$, because of its dependence on the cosmological factor $dt/dz$, which in turns depends on the inverse of $H_{0}$. Although differences are small, such an expectation is confirmed at the lower distance (see left panel of Fig.\,\ref{fig-gzk-nu}). On the other hand, the number density of background photons increases with distance, as well as their energy: such a behavior tends to dominate the previous one for distance above $\approx1$~Gpc. The differences in the distance at which GZK neutrinos are produced, have a direct impact on the energy spectrum of protons ($\Phi_{p}(E)$) and neutrinos ($\Phi_{\nu}(E)$) observed at Earth. In the case of neutrinos, $\Phi_{\nu}(E)$ is shown in the right panel of Fig.\,\ref{fig-gzk-nu}. 

\begin{figure}[!t]
	\centering
	  \includegraphics[width=11.0cm]{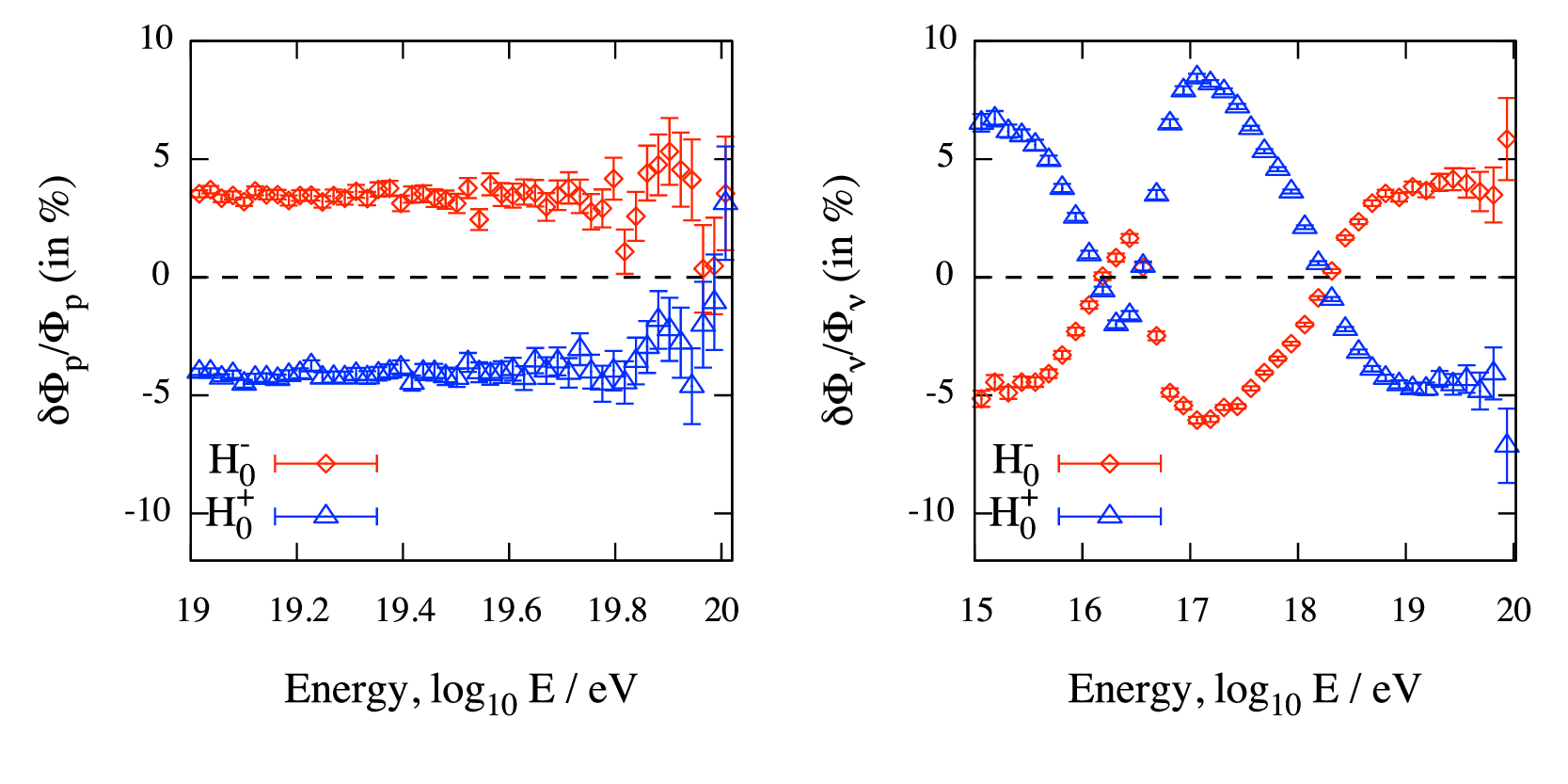}
	\caption{Same simulation setup as in Fig.\,\ref{fig-gzk-nu}: relative difference between fluxes at Earth corresponding to $H_{0}^{+}$ and $H_{0}^{-}$, with respect to $H_{0}$, in the case of protons (left panel) and neutrinos (right panel).}
\label{fig-gzk-flux}
\end{figure}

The relative differences between the flux of protons and neutrinos corresponding to $H_{0}^{\pm}$ and $H_{0}$, are shown in Fig.\,\ref{fig-gzk-flux}. The effect of varying the Hubble parameter is a simple diagonal shift in the flux. Let $\delta\Phi(E)/\Phi(E)$ indicate the relative difference of the energy spectrum corresponding to scenarios with $H_{0}^{\pm}$ with respect to scenario with $H_{0}$. In the case of protons, where a simple power-law flux is expected up to $40-50$~EeV, such a shift induced by $H_{0}^{\pm}$ should produce a rather constant $\delta\Phi_{p}(E)/\Phi_{p}(E)$ versus energy. Such an expectation is confirmed by the result shown in the left panel of Fig.\,\ref{fig-gzk-flux}, where constant differences of the order of 4-5\% are found at any energy below $\approx$60~EeV. We argue that such differences can be quantitatively explained by the different energy-loss rates corresponding to the values of $H_{0}^{\pm}$ and $H_{0}$. In fact, the energy-loss rate is proportional to the cosmological factor $-dt/dz$, and, by consequence, to the inverse of the Hubble parameter at the present time. It can be shown that the diffuse spectrum reflects such a proportionality to the inverse of $H_{0}$ because of its direct dependence on the cosmological factor (see, for instance, Ref.\,\cite{berezinsky2006astrophysical}). Hence, the expected difference in the flux of UHE protons approximately reduces to $(H_{0}-H_{0}^{\pm})/H_{0}^{\pm}$, providing a $\approx5\%$ alteration of the flux in the case of $H_{0}^{-}$ and a $\approx-5\%$ alteration of the flux in the case of $H_{0}^{+}$, in good agreement with our finding.

In the case of neutrinos, the more complicated dependence on energy of $\Phi_{\nu}$ is expected to produce non-constant relative differences $\delta\Phi_{\nu}(E)/\Phi_{\nu}(E)$. The results shown in the right panel of Fig.\,\ref{fig-gzk-flux} confirms such an expectation: relative differences range from 0\% to $\approx$10\%. Such a result suggests that experimental evidences of this deviation should be explored below $10^{16}$~eV, around $10^{17}$~eV or above $10^{19}$~eV. However, it is worth remarking that a method to exploit such a deviation to probe the Hubble parameter is still under investigation and it will be the subject of successive studies.

\subsection{Application \#2: clustering}

\begin{figure}[!t]
	\centering
	  \includegraphics[width=11.0cm]{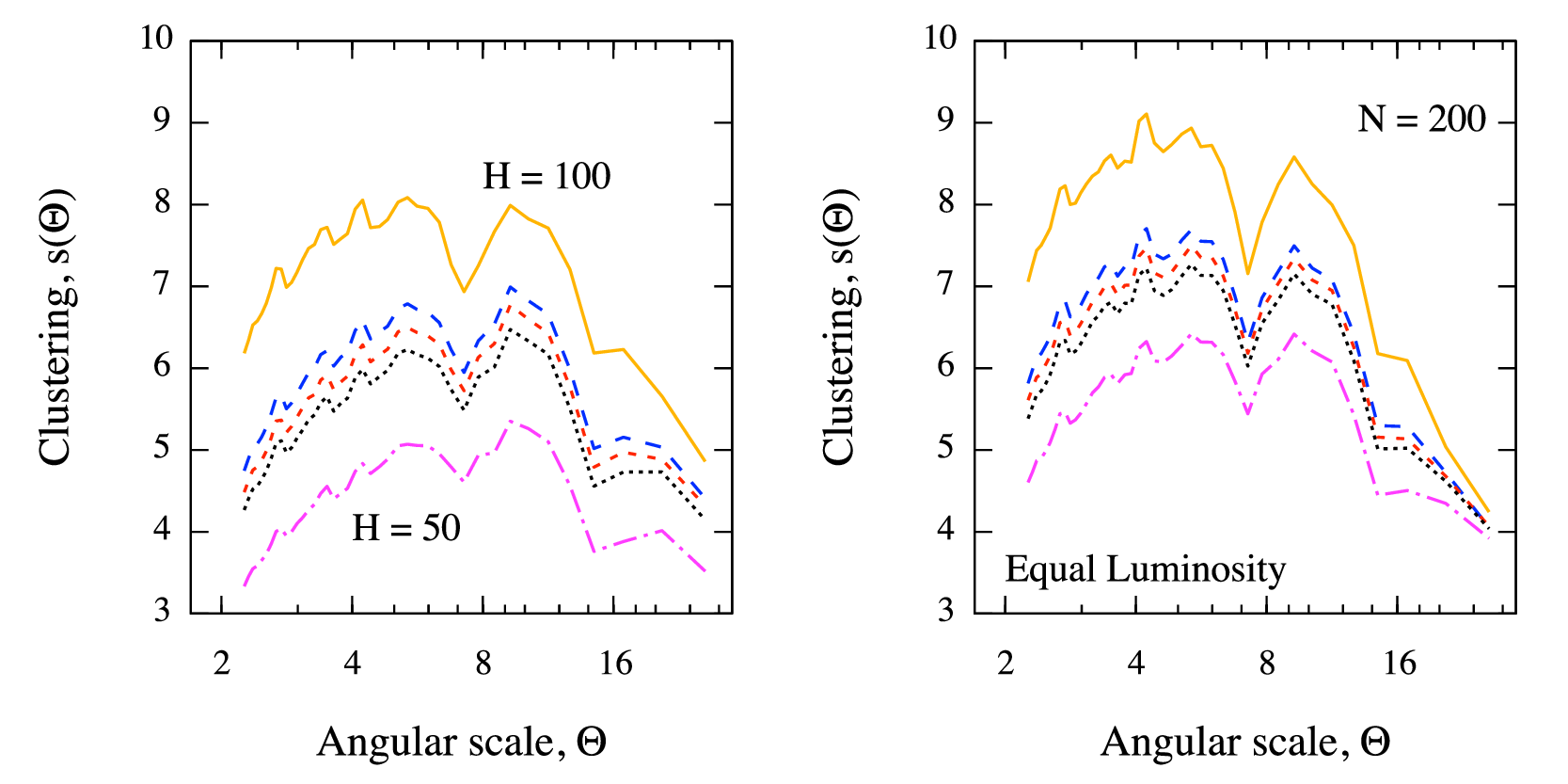}
	\caption{Expected clustering signal, as a function of the angular scale (in deg units), from a sky of $N=200$ protons with $E\geq 100$~EeV (in the field of view of Pierre Auger Observatory) and for values of the Hubble parameter considered in Fig.\,\ref{fig-wgzkH}. Sources of 44\% of events are AGN within $z=0.047$ in the SWIFT-BAT 58-months catalog, whereas the remaining 56\% of events are isotropically distributed. Scenarios with intrinsic luminosity taken into account (left panel) and not taken into account (right panel) are considered. The signal at each angular scale is obtained by averaging over $10^{4}$ Monte Carlo realizations.}
\label{fig-clustering-first}

	  \includegraphics[width=11.0cm]{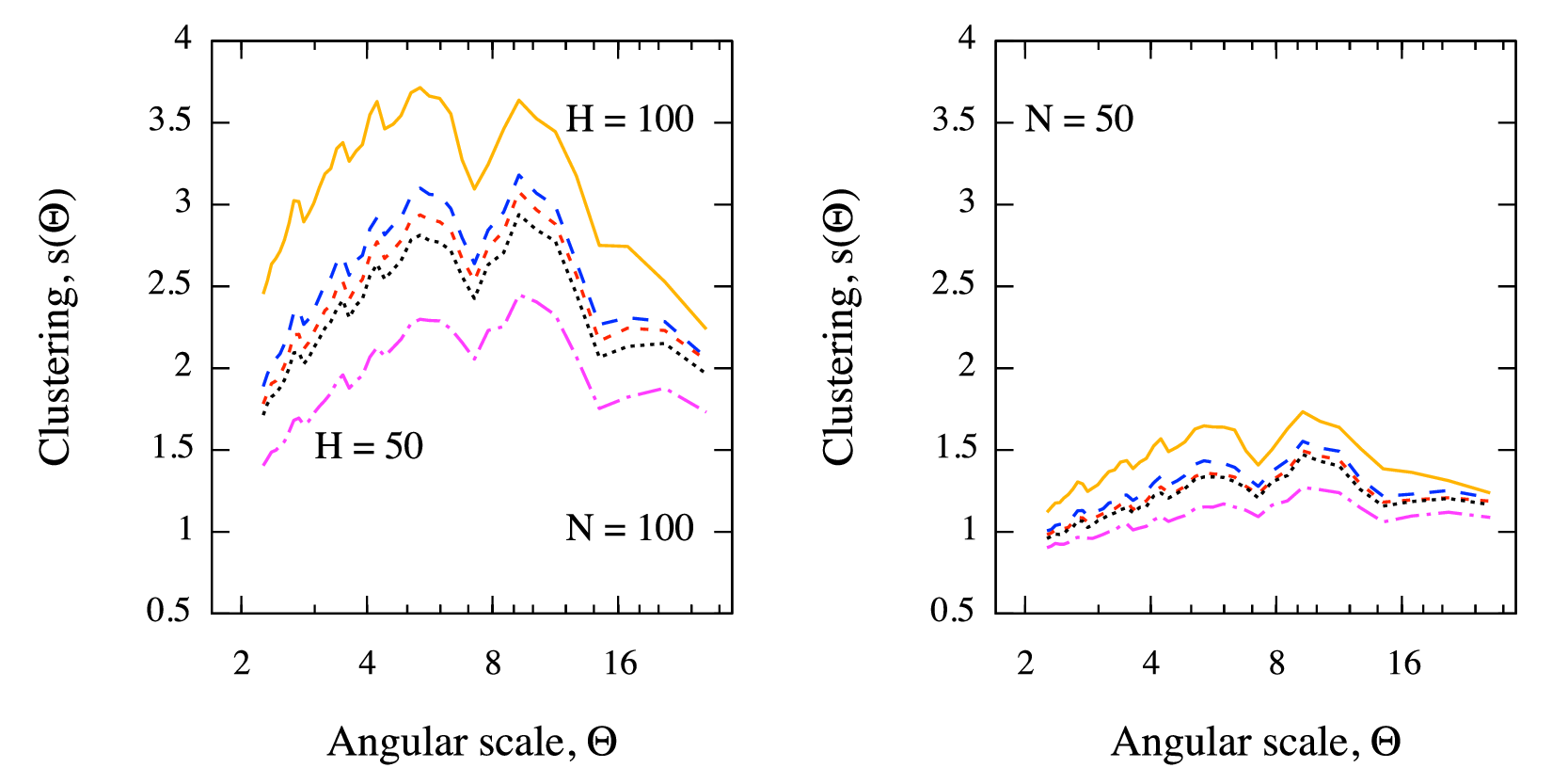}
	\caption{Same as in Fig.\,\ref{fig-clustering-first}, for scenarios where intrinsic luminosity of AGN is taken into account. The clustering for $N=100$ (left panel) and $N=50$ (right panel) is considered.}
\label{fig-clustering-second}
\end{figure}

Motivated by the recent correlation between the arrival directions of UHECRs detected with the Pierre Auger Observatory and AGN \cite{auger2010correlation}, as a second practical application, we consider the distribution of sources corresponding to the position of AGN in the nearby Universe (up to $z=0.047$), reported in the SWIFT-BAT 58-months catalog \cite{swift2010}. In particular, we consider two scenarios: i) we assume equal intrinsic luminosity for all AGN and ii) the intrinsic luminosity of each AGN is taken into account. If $\mathcal{L}$ indicates luminosity of an AGN, and by assuming no source evolution, the probability to get an event from such a source is proportional to $\mathcal{L}~z^{-2}\omega_{\text{GZK}}(z,E_{\text{thr}})$. Thus, protons are then propagated in a $\Lambda$CDM Universe until they reach the Earth. We consider only UHECRs with energy above $100$~EeV and with arrival direction lying in the field of view of the Pierre Auger Observatory, whose non-uniform exposure is taken into account, as well as its angular uncertainty of $0.8^{\circ}$. The effect of EMF is also taken into account, smearing the direction around the source by sampling a Fisher-von Mises distribution, i.e. the Gaussian counterpart on the sphere. The spreading angle is given by Eq.\,(\ref{def-sigl-theta}) in the case of r.m.s. strength $B_{\text{rms}}=2$~nG and correlation length $\ell=1$~Mpc, according to the most recent upper bounds \cite{trivedi2010primordial}. Additionally, according to the result reported by Pierre Auger Collaboration in the case of the SWIFT-BAT 58-months catalog, the 56\% of events in the simulated sky are isotropically distributed \cite{auger2010correlation}.

We investigate the clustering signal averaged over several Monte Carlo realizations ($10^{4}$ for each astrophysical scenario), by mean of the novel multiscale autocorrelation function (MAF) \cite{dedomenico2011MAF}, versus the angular scale. Such a function involves the equal-area binning of a spherical region of the sky: the number $N$ of bins defines the angular scale $\Theta$ of the analysis. Let $\psi_{k}(\Theta)$ be the fraction of points from the data set and $\overline{\psi}_{k}(\Theta)$ be the expected isotropic fraction in the bin $\mathcal{B}_{k}$. The Kullback-Leibler divergence \cite{kullback1951information}
\begin{eqnarray}
\label{def-A}
A(\Theta) = \mathcal{D}_{\text{KL}}\( \psi(\Theta) || \overline{\psi}(\Theta)\) = \sum_{k=1}^{N} \psi_{k}(\Theta)\log\frac{\psi_{k}(\Theta)}{\overline{\psi}_{k}(\Theta)}
\end{eqnarray}
quantifies the departure from an isotropic distribution at the scale $\Theta$. If $A_{\text{data}}(\Theta)$ and $A_{\text{iso}}(\Theta)$ refer, respectively, to the data and to an isotropic realization with the same number of events, the MAF estimator is defined as the standardized deviation from isotropy:
\begin{eqnarray}
\label{def-s}
s(\Theta)=\frac{\left|A_{\text{data}}(\Theta)-\left\langle A_{\text{iso}}(\Theta) \right\rangle\right|}{\sigma_{A_{\text{iso}}}(\Theta)},
\end{eqnarray}
where $\left\langle A_{\text{iso}}(\Theta) \right\rangle$ and $\sigma_{A_{\text{iso}}}(\Theta)$ are the sample mean and the sample standard deviation, respectively, estimated from several isotropic realizations of the data. The MAF estimator is not biased against the null hypothesis $\mathcal{H}_{0}$ of an underlying isotropic distribution for the data and the estimated chance probability depends on $\Theta$. Indeed, $s(\Theta)$ follows a half-Gaussian distribution with zero mean and unitary variance, and the probability to obtain a maximum value of $s(\Theta)$, at any angular scale $\Theta$, greater or equal than a given value $s^{\star}$ is
\begin{eqnarray}
\label{def-p}
p\(s^{\star}\)=1-\exp\[ -\exp\(\frac{s^{\star}-1.743}{0.470}\) \],
\end{eqnarray}
providing an analytical expression for the properly penalized chance probability, independently on the value of the angular scale $\Theta$ and on the data size. It is worth remarking that the angular scale where the chance probability is minimum (i.e. where $s(\Theta)$ is maximum) turns to be the most relevant clustering scale and that, for each $\Theta$, the value of $s(\Theta)$ is an estimation of the amount of clustering at that scale \cite{dedomenico2011MAF}.

In this study, different values of the parameter $H_{0}$ are considered, as well as an increasing number of events in the sky. The results are shown in Fig.\,\ref{fig-clustering-first} and \ref{fig-clustering-second}, for different astrophysical scenarios and angular scales (in deg units). In Fig.\,\ref{fig-clustering-first} we consider the cases where intrinsic luminosity of AGN is taken into account (left panel) and not taken into account (right panel), for skies of $N=200$ protons. In Fig.\,\ref{fig-clustering-second} we focus on scenarios where intrinsic luminosity is accounted for, and vary the number of protons, to put in evidence the impact of the statistics on the clustering signal. It is evident that, for a fixed number of events, the clustering signal increases for increasing values of $H_{0}$, whereas it decreases for decreasing number of events, as expected\footnotemark\footnotetext{The statistical power of the method increases with the number of events.}. The dependence of the clustering signal on the value of the Hubble parameter at the present time can be understood in terms of the surviving probability defined by Eq.\,(\ref{def-wgzk}) and (\ref{def-wgzk-flux}). In fact, we have previously shown that changes in the value of $H_{0}$ have a non-negligible impact on the surviving flux $\Omega_{\text{GZK}}(z)$. Such an impact is reflected in the weight function $\omega_{\text{GZK}}(z)$ adopted in our simulations for the probability to get an event from a source. For a fixed distance, the probability to reach the Earth for UHE protons propagating in a $\Lambda$CDM Universe increases for increasing values of $H_{0}$, as clearly deducible from Fig.\,\ref{fig-wgzkH}. Such a behavior favors the clustering around nearby sources, increasing the signal at any angular scale.

\begin{figure}[!t]
	\centering
	  \includegraphics[width=11.0cm]{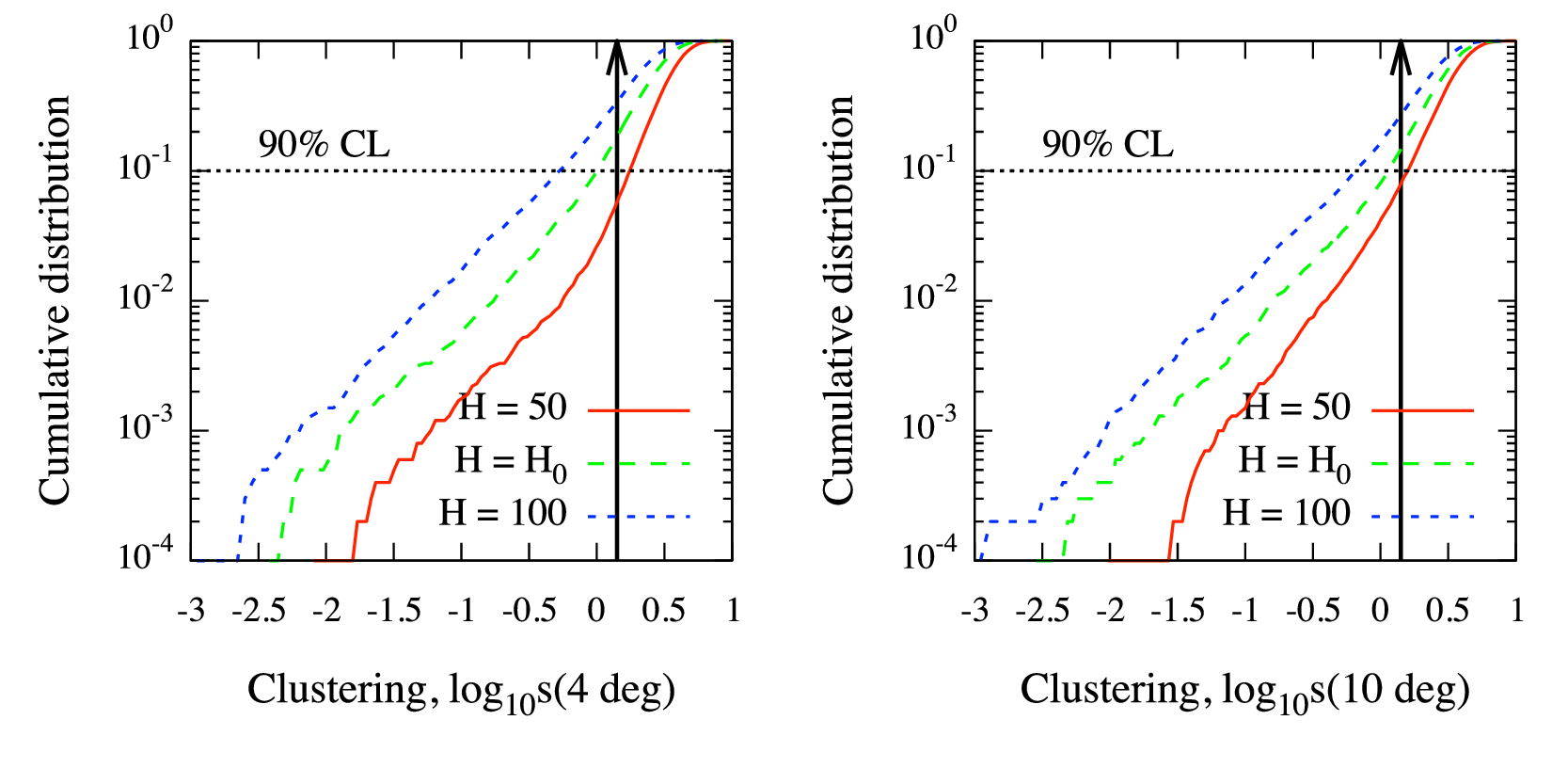}
	\caption{Same scenario as in Fig.\,\ref{fig-clustering-second}, with $N=100$ events above $E_{\text{thr}}=100$~EeV. It is shown the cumulative distribution of the clustering function $s(\Theta)$ in the case with $\Theta=4^{\circ}$ (left panel) and $\Theta=10^{\circ}$ (right panel), for different values of the Hubble parameter. For each scenario, $10^{4}$ Monte Carlo realizations have been considered. As an illustrative example, we assume that the clustering coefficient obtained from the data is 1.4 (solid vertical arrow): in both panels, we show a test with 90\% confidence level (dotted line) against such a value. See the text for further details.}
\label{fig-clustering-cdf}
\end{figure}

A direct comparison, as a function of the angular scale, between the clustering signal $s(\Theta)$ obtained from the data and that one obtained from simulations, for different values of $H_{0}$, represents the principal tool for probing the Hubble parameter with our clustering approach. The procedure is outlined in the following. 

Let us assume that observations provide a number $N$ of events above a certain energy threshold $E_{\text{thr}}$. We consider an astrophysical scenario with a certain value of the Hubble parameter $H$, by using a catalog of real candidate sources producing a fraction $f_{p}$ of UHECRs. The remaining fraction $1-f_{p}$ of particles is distributed isotropically in the sky. Hence, we simulate a large number of skies with $N$ events in the field of view of the observatory (taken into account the non-uniform exposure, if required), and we estimate the multiscale clustering $s(\Theta)$ from both the data and the simulations. For a fixed angular scale $\Theta_{0}$, the statistical test can be performed by comparing the distribution of $s_{\text{sim}}(\Theta_{0}; H)$, obtained from the model, against the value $s_{\text{obs}}(\Theta_{0})$, obtained from the data. For instance, either one- or two-tailed test can be adopted to accept or reject null hypothesis that the observation is compatible with the expectation. In order to gain full advantage from information provided by the multiscale approach, such a procedure can be repeated for any value of the angular scale, and the results can be combined with the Fisher's method, its extensions, or any other statistical method. 

As an illustrative example of our procedure, in Fig.\,\ref{fig-clustering-cdf} we show the case of a one-tailed statistical test with 90\% CL (indicated by the dotted horizontal line). In absence of values of $s_{\text{obs}}$ obtained from real data, we assume, for instance, that $s_{\text{obs}}=1.4$ (indicated by the solid vertical arrow in the figure). Moreover, we consider the same astrophysical scenario adopted to obtain the results shown in Fig.\,\ref{fig-clustering-second}, with $N=100$ events above $E_{\text{thr}}=100$~EeV. In particular, we show the cumulative distributions of $s_{\text{sim}}(\Theta_{0}; H)$ for three different values of the Hubble parameter, namely 50, 70.4 and 100~km/s$^{-1}$/Mpc, and for two different values of the angular scale, namely $4^{\circ}$ and $10^{\circ}$. For such a test, the null hypothesis can not be rejected if the intersection between the line corresponding to $s_{\text{obs}}$ and the curve corresponding to a scenario with a certain Hubble parameter, lies below the CL line.

Hence, in both examples, it is evident that the null hypothesis should be rejected in the cases corresponding to $H=70.4$ and $H=100$, whereas it can not be rejected for $H=50$ at 90\% CL.

Hence, such a study suggests that the observation of 100 protons with energy above $100$~EeV should be sufficient to probe the Hubble parameter. The large statistics required for this investigation can be attained in few years of activity by future arrays much larger than the Pierre Auger Observatory and with larger exposure, as, for instance, JEM-EUSO. However, it is worth remarking that we have considered only the most conservative case, with events above $100$~EeV. In Fig.\,\ref{fig-gzk-rel-diff} is shown that an energy threshold of 60~EeV is sufficient enough to observe significant differences in the flux, and, indirectly, on the clustering of protons. At such energy threshold, we already expect that current experiments as the Pierre Auger Observatory will collect more than 200 UHECRs within a few years, providing the statistics required for this probe.

However, it is worth remarking that such a result depends on the assumptions about the sources of UHECRs and, by consequence, about their distribution in the nearby Universe. In fact, the clustering signal measured in the arrival direction distribution of UHECRs is sensitive to the intrinsic clustering of sources and their number density \cite{dedomenico2011bounds}. Moreover, the (still unknown) composition of UHECRs plays a significant role in the formation of clusters of particles. In fact, both extragalactic and galactic magnetic fields have a negligible impact on the deflection of UHE protons. Conversely, magnetic fields are expected to bend significantly the trajectories of UHE heavier nuclei, diluting the clustering signal at the smallest angular scales and altering the signal at the largest ones. In any case, our procedure is robust against the ignorance about the intervening magnetic fields and the composition of UHECRs, if the fraction of protons (adopted as input to the simulations) is estimated from the observations, as for instance in Ref.\,\cite{auger2010correlation}, and if the remaining fraction of UHECRs is prevalently composed by heavy nuclei. Hence, it is clear that without a definitive knowledge of the UHECR source population, the inferred value of the Hubble parameter at the present time would depend on the underlying assumptions about the candidate sources adopted for the study.


\section{Conclusion}

The GZK effect plays a fundamental role in the search of sources of UHECRs. Within the present work we have investigated the influence of cosmology on the GZK horizon of extragalactic UHE protons, with energy ranging from $50$ to $100$ EeV. By considering very different models of the Universe, from flat to curved ones,  we have shown that significant differences among cosmological models appear to be important above the GZK horizon, where the surviving probability for the protons is very small. Moreover, we have investigated the impact of uncertainty in the Hubble parameter at the present time, according to the $\Lambda$CDM model of the Universe and to the experimental constraints obtained from recent WMAP observations. Our results suggest the existence of non-negligible differences between the estimated values of the GZK horizon in Universes with Hubble parameter $H_{0}=70.4+3\sigma$ and $H_{0}=70.4-3\sigma$, respectively. However, our numerical results show that such differences should have a small impact on studies involving distances below 250~Mpc, as for instance the recent correlation analyses between observed data and the distribution of nearby active galactic nuclei (AGN) reported by the Pierre Auger and the HiRes collaborations \cite{auger2007correlation,auger2008correlation,auger2010correlation,abbasi2008search}. Finally, we have shown that current and future experiments could probe the value of the Hubble parameter by measuring the flux of GZK neutrinos at Earth or the clustering signal of UHE protons.

Authors thank C. Dobrigkeit, P. L. Ghia and C. Inserra for critically reading the manuscript and for useful comments. M.D.D also acknowledges H. Lyberis for suggestions and stimulating discussions. Anonymous referees are also acknowledged for their fruitful comments and suggestions.


\section*{References}
\addcontentsline{toc}{section}{References} 
\begin{small}
\bibliographystyle{unsrt} 
\bibliography{draft}
\end{small}

\end{document}